\documentclass[fleqn,usenatbib]{mnras}

\usepackage{newtxtext,newtxmath}

\usepackage[T1]{fontenc}
\usepackage{ae,aecompl}

\usepackage{myaasmacros}
\usepackage{graphicx}	
\usepackage{amsmath}	
\usepackage{booktabs,tabularx,graphicx,amsmath,makecell,hhline,enumitem,gensymb,multirow,ulem,pifont} 
\usepackage[dvipsnames]{xcolor}
\definecolor{Halo1}{HTML}{305FEA}
\definecolor{Halo2}{HTML}{00A982}
\definecolor{Halo3}{HTML}{7F0579}
\definecolor{Halo4}{HTML}{00D1D4}
\definecolor{Halo5}{HTML}{EF1414}
\definecolor{GM}{HTML}{EAA209}
\definecolor{GM2}{HTML}{000000}

\title[Two routes to dark matter core formation in ultra-faint dwarfs]{EDGE: Two routes to dark matter core formation in ultra-faint dwarfs}

\author[Matthew D. A. Orkney et al.] 
{Matthew D. A. Orkney$^{1}$,
Justin I. Read$^{1}$, 
Martin P. Rey$^{2}$,
Imran Nasim$^{1}$,
Andrew Pontzen$^{2}$, 
Oscar Agertz$^{3}$,
\newauthor
Stacy Y. Kim$^{1}$,
Maxime Delorme$^{4}$,
Walter Dehnen$^{5,6,7}$
\\
$^1$Department of Physics, University of Surrey, Guildford, GU2 7XH, United Kingdom\\
$^2$Department of Physics and Astronomy, University College London, London WC1E 6BT, UK\\
$^3$Lund Observatory, Department of Astronomy and Theoretical Physics, Lund University, Box 43, SE-221 00 Lund, Sweden\\
$^4$ Département d'Astrophysique/AIM, CEA/IRFU, CNRS/INSU, Université Paris-Saclay, 91191 Gif-Sur-Yvette, France\\
$^5$ Astronomisches Recheninstitut, Zentrum f{\"u}r Astronomie der Universit{\"a}t Heidelberg, M{\"o}nchhofstra\ss{}e. 12-14, 69120, Heidelberg, Germany\\
$^6$ Universit{\"a}ts-Sternwarte M{\"u}nchen, Scheinerstra\ss{}e 1, 81679, M{\"u}nchen, Germany\\
$^7$ School for Physics and Astronomy, University of Leicester, University Road, LE1 7RH, UK
}
\date{Submitted to MNRAS}

\pubyear{2020}

\begin{document}
\label{firstpage}
\pagerange{\pageref{firstpage}--\pageref{lastpage}}
\maketitle

\begin{abstract}
In the standard Lambda cold dark matter paradigm, pure dark matter simulations predict dwarf galaxies should inhabit dark matter haloes with a centrally diverging density `cusp'. This is in conflict with observations that typically favour a constant density `core'. We investigate this `cusp-core problem' in `ultra-faint' dwarf galaxies simulated as part of the `Engineering Dwarfs at Galaxy formation's Edge' (EDGE) project. We find, similarly to previous work, that gravitational potential fluctuations within the central region of the simulated dwarfs kinematically heat the dark matter particles, lowering the dwarfs' central dark matter density. However, these fluctuations are not exclusively caused by gas inflow/outflow, but also by impulsive heating from minor mergers. We use the genetic modification approach on one of our dwarf's initial conditions to show how a delayed assembly history leads to more late minor mergers and, correspondingly, more dark matter heating. This provides a mechanism by which even ultra-faint dwarfs ($M_* < 10^5\,\text{M}_{\odot}$), in which star formation was fully quenched at high redshift, can have their central dark matter density lowered over time. In contrast, we find that late major mergers can regenerate a central dark matter cusp, if the merging galaxy had sufficiently little star formation. The combination of these effects leads us to predict significant stochasticity in the central dark matter density slopes of the smallest dwarfs, driven by their unique star formation and mass assembly histories.
\end{abstract}




\section{Introduction} \label{sec:intro}

The $\Lambda$CDM paradigm presents us with a Universe that has an energy budget dominated by dark matter and dark energy, and in which galaxies are assembled through successive hierarchical mergers \citep{1978MNRAS.183..341W}. It has proven to be extremely successful in predicting the formation of cosmic structure on large scales \citep{2006Natur.440.1137S, 2006ApJ...648L.109C, 2006PhRvD..74l3507T, 2013AJ....145...10D, 2014MNRAS.439.2515O, 2014A&A...571A..16P, 2016MNRAS.456.2301W}. 
However, disagreements between theory and observation endure at (sub-)galactic scales that have become collectively known as `small scale puzzles' \citep[e.g.][]{2017ARA&A..55..343B}. 

The oldest, and perhaps most challenging, of the small scale puzzles is the `cusp-core problem' (CC) \citep[e.g.][]{1994ApJ...427L...1F, 1994Natur.370..629M, 2017MNRAS.467.2019R}. Pure dark matter structure formation simulations in $\Lambda$CDM predict a self-similar radial dark matter density profile -- the NFW profile \citep{navarro1997universal}. This scales as $\rho \propto r^{-1}$ within a scale radius $r_s$, referred to as a `cusp' due to its divergence towards the origin. By contrast, observations of rotation curves in dwarf galaxies appear to favour instead a constant inner dark matter density, referred to as a `core' \citep[e.g.][]{1988ApJ...332L..33C, 1994ApJ...427L...1F, 1994Natur.370..629M, 2001AJ....122.2381M,2017MNRAS.467.2019R}.

Many solutions to the CC have been proposed in the literature to date. Firstly, observations could have been misinterpreted due to incorrect modelling assumptions. Typical assumptions include spherical symmetry, circular orbits and dynamical pseudo-equilibrium, all of which can be reasonably questioned \citep[e.g.][]{2011MNRAS.414.3617K, 2016MNRAS.460.3610O,2016MNRAS.462.3628R}. Secondly, the assumed underlying dark matter model could be incorrect. Alternatives such as warm dark matter \citep[e.g.][]{2000PhRvD..62f3511H, 2001ApJ...556...93B, 2001ApJ...559..516A}, self-interacting dark matter \citep[e.g.][]{2000PhRvL..84.3760S,2018PhR...730....1T} or ultra-light dark matter \citep[e.g.][]{2014NatPh..10..496S,2020arXiv200503254F} all predict a lower dark matter density at the centres of dwarf galaxies whilst retaining the predictions of $\Lambda$CDM on larger scales. However, in recent years a third class of solution has been gaining traction.

The CC problem originates from a comparison of pure dark matter simulations --- that do not model stars or gas (hereafter baryons) --- with observations. This opens up the possibility that purely gravitational interactions between dark matter particles and baryons could act to push dark matter out from the centres of dwarf galaxies, transforming a cusp into a core. Three main mechanisms have been proposed to date:
\setlist{leftmargin=15pt,labelindent=15pt}
\setlist[enumerate]{wide=0pt,widest=99,leftmargin=2\parindent,labelsep=*,align=left}
\begin{enumerate}
    \item Dynamical friction from infalling dense clumps \citep[e.g.][]{2001ApJ...560..636E, 2004MNRAS.353..829M, 2009ApJ...702.1250R, 2010ApJ...725.1707G, 2015MNRAS.446.1820N}. These clumps impart energy and angular momentum to the dark matter halo, causing it to expand.
    \item Dynamical friction from a central stellar or gaseous bar that acts similarly to infalling clumps, kinematically `heating' the background dark matter halo. \citep[e.g.][]{2007MNRAS.375..460W}.
    \item A fluctuating gravitational potential driven by gas inflow/outflow due to cooling, stellar winds and supernovae. This causes dark matter particle orbits to slowly migrate outwards \citep[e.g.][]{1996MNRAS.283L..72N, 2005MNRAS.356..107R, 2008Sci...319..174M, 2012MNRAS.421.3464P}.
\end{enumerate}
All three mechanisms owe, ultimately, to a time-varying gravitational potential. This allows dark matter particles to exchange orbital energy both with one another and with the stars and gas in the galaxy. What differs is only the physical mechanism that drives the time-dependent gravitational field. In principle, all three mechanisms can act in tandem as galaxies form and evolve. \par

Despite this diversity of mechanisms in the literature, to date high resolution galaxy formation simulations have typically favoured mechanism (iii) at the scale of dwarf galaxies \citep{2012MNRAS.421.3464P, 2014Natur.506..171P, 2013MNRAS.429.3068T, 2014MNRAS.441.2986D, 2015MNRAS.454.2092O, 2016MNRAS.461.2658D}. Once gas is allowed to cool $(T<10^4$\,K) and reach high density ($\rho > 10$\,atoms/cc; e.g. \citealt{2012MNRAS.421.3464P,2016MNRAS.461.2658D}), these simulations find that gas flows drive repeated fluctuations in the central galaxy mass of amplitude $10-20$\% over a period less than the local dynamical time. Such fluctuations gradually lower the inner dark matter density on the scale of the stellar half mass radius, $R_{1/2}$, transforming a dark matter cusp to a core \citep[e.g.][]{2015MNRAS.454.2981C,2016MNRAS.459.2573R}. There is mounting observational evidence that this process occurs in real dwarf galaxies (e.g. \citealt{2014MNRAS.441.2717K, 2016ApJ...820..131E, 2017MNRAS.466...88S, 2019MNRAS.484.1401R, 2019ApJ...880...54H}; but see also \citealt{2019MNRAS.486.4790B,2019MNRAS.482..821O,2020arXiv201109482G}).

While dark matter heating may solve the cusp core problem in isolated gas rich dwarfs, a new puzzle has recently presented itself: there is a growing body of evidence for small dark matter cores even within `ultra-faint' dwarf galaxies, typically defined to have stellar masses $M*<10^5$\,M${_\odot}$ \citep[e.g.][]{2017ApJ...844...64A, 2018MNRAS.476.3124C, 2018MNRAS.478.3879S, 2020MNRAS.tmp.3382M, 2020arXiv201200043S}. Several papers have suggested that galaxies with so few stars have insufficient energy from stellar feedback to carve out a dark matter core of size $0.5-1$\,kpc (e.g. \citealt{2012ApJ...759L..42P, 2013ApJ...766...56M, 2015MNRAS.454.2092O, 2016MNRAS.456.3542T}). However, dark matter cores form on the scale of the half mass radius, $R_{1/2}$ \citep{2016MNRAS.459.2573R}, which can be as small as $30-300$\,pc in ultra-faint dwarfs \citep[e.g.][]{2019ARA&A..57..375S}. Such small cores form much more rapidly and require significantly less energy, raising the possibility that dark matter core formation could proceed `all the way down' to even the smallest dwarfs \citep[e.g.][]{2016MNRAS.459.2573R,2018MNRAS.476.3124C}. Furthermore, such small cores remain dynamically important by construction since they exist precisely where the stars and gas do -- i.e. precisely where we can hope to measure the inner dark matter potential.

In this paper, we use a suite of high resolution cosmological zoom simulations from the Engineering Dwarfs at Galaxy formation's Edge (EDGE) project \citep{2019ApJ...886L...3R, 2020MNRAS.491.1656A, 2020MNRAS.497.1508R, 2020MNRAS.tmp.3431P} to explore whether dark matter core formation can proceed even in the very smallest dwarf galaxies. Our simulations model galaxies over the mass range $M_{200{\rm c}} \sim 1-4 \times 10^9$\,M$_\odot$, consistent with ultra-faint dwarfs, and reach a spatial resolution of $\sim 3$\,pc, sufficient to resolve even very small dark matter cores.

This paper is organised as follows. In Section~\ref{simulation methods}, we describe the EDGE simulations and our numerical methods. In Section~\ref{overview}, we present a first visual impression of the EDGE simulation suite. In Section~\ref{gasflows}, we investigate how gas flows driven by bursty star formation drive early-time core formation in our simulations.
In Section~\ref{mergers}, we show that minor mergers can also drive the formation of cores in EDGE and we validate the robustness of our results using the $N$-body code {\sc griffin}.
In Section~\ref{core destruction}, we show how late major mergers can reintroduce a dark matter cusp. In Section~\ref{discussion}, we discuss the implications of our results for dark matter cusps and cores in the smallest dwarf galaxies. Finally, in Section~\ref{conclusion} we present our conclusions.

\section{Method}
\subsection{Simulations} \label{simulation methods}

The EDGE project is described in detail in \citet{2020MNRAS.491.1656A}. Here, we briefly summarise the key points. We start with a $512^3$ resolution cosmological dark matter simulation of a $50\,$Mpc void region (Figure \ref{fig:void}). All simulations assume cosmological parameters $\Omega_m=0.309$, $\Omega_\Lambda=0.691$, $\Omega_b=0.045$ and $H_0=67.77\,\text{km\,s}^{-1}\,\text{Mpc}^{-1}$, in line with data from the PLANCK satellite \citep{2014A&A...571A..16P}.

We draw a selection of target haloes from the void volume, chosen from a range in halo mass of $10^9<M/\text{M}_{\odot}<5\times10^{9}$. These target haloes are resimulated following the zoom simulation technique \citep{1993ApJ...412..455K, 2014MNRAS.437.1894O} with the Adaptive Mesh Refinement (AMR) code {\sc ramses} \citep{2002A&A...385..337T}. This grants us a highly resolved target galaxy within its lower-resolution wider cosmological context. The velocity in our initial conditions is then adjusted to match the velocity of the target halo, which reduces the impact of numerical diffusion effects \citep[see][]{2020MNRAS.tmp.3431P}. \par

Key details of our {\sc ramses} simulations are presented in Table \ref{tab:sims}. Each simulation is run over the redshift range $99 \geq z \geq 0$, with a minimum of 100 outputs spaced linearly with the scalefactor, $a$. The contamination fraction, defined as the fraction of lower resolution dark matter particles within the virial radius, is never greater than $2\times10^{-5}$ in any of our simulations. This is relevant for the impact of numerical relaxation, which we discuss further in Appendix \ref{appendix:a}. Our simulations are run at a resolution in which the dark matter particle mass approaches $100\,$M$_{\odot}$ in the high resolution Lagrangian region of the target galaxy, with a spatial resolution $\sim 3\,$pc in the most resolved zoom regions. At this resolution, the momentum injection of individual supernovae into the interstellar medium can be accurately resolved \citep{2015MNRAS.451.2900K}, avoiding the need for delayed cooling, inflated SNe energies, or sub-grid wind models \citep[e.g.][]{2016MNRAS.459.2573R,2020MNRAS.491.1656A}. \par

Star formation is described with a Schmidt law \citep{1959ApJ...129..243S, 1998ApJ...498..541K} in gas cells that meet certain density and temperature requirements:
\begin{equation}
\label{schmidt.eq}
    \Dot{\rho_*} = \epsilon_{\rm ff} \frac{\rho_g}{t_{\rm ff}} \text{ for } \rho_g > \rho_\star \text{ and } T_g < T_\star,
\end{equation}
where $\rho_\star = 300\,m_{\rm proton}\,\text{cm}^{-3}$ and $T_\star = 100\,\text{K}$. Here, $\Dot{\rho_*}$ is the star formation rate density in a gas cell, $\rho_{g}$ is the density per gas cell, $t_{\rm ff} = \sqrt{3\pi/32G\rho_g}$ is the
local free-fall time of the gas, and $\epsilon_{\rm ff}$ is the star formation efficiency per free-fall time which is set to 10\% in line with arguments from \citet{2019MNRAS.486.5482G}. Each stellar particle is initialised at $300\,\text{M}_{\odot}$ and is representative of a single-age stellar population (SSP) described by a Chabrier initial mass function (IMF) \citep{2003PASP..115..763C}. Stellar feedback from both Type II and Ia supernovae are included, and stellar winds from massive and asymptotic giant branch (AGB) stars (see \citealt{2013ApJ...770...25A, 2015ApJ...804...18A, 2020MNRAS.491.1656A} for details). \par

The epoch of reionisation is modelled as a time-dependent uniform UV background around $z=8.5$, as in the public release of {\sc ramses} \citep{1996ApJ...461...20H}. The exact implementation is discussed further in \citet{2020MNRAS.497.1508R}, and is consistent with a late reionisation expected for a cosmic void \citep[e.g.][]{2020MNRAS.491.1736K}. \par 

\begin{table*}
\caption{Details of all simulations. The simulation labels denote different reference haloes selected from the initial void simulation. From left to right the columns are: the simulation label, the physics scheme employed, the mass resolution, the halo mass ($M_{200 \rm c}$) at $z=0$, total stellar mass within the virial radius ($r_{200 \rm c}$) at $z=0$, the virial radius at $z=0$, the projected half light radius at $z=0$, and the V-band magnitude at $z=0$. The simulations are ordered by the $M_{200 \rm c}$ mass of the full physics simulations.}
\begin{tabular}{lcccccccc}
 \hline\hline
 \textbf{Name} & \textbf{Physics} & \textbf{Resolution} & \textbf{$M_{200 \rm c}$ [$\text{M}_{\odot}$]} & \textbf{$r_{200 \rm c}$ [kpc]} & \textbf{$M_{*}$ [M$_{\odot}$]} & \textbf{$R_{\rm half}$ [pc]} & \textbf{$M_V$ (mag)} \\
  & & \textbf{[$m_{\rm DM},m_{\rm gas},m_{\rm *}]/\text{M}_{\odot}$} & & & & (projected) & \\
 \hline\hline
 \textcolor{Halo1}{Halo1445 DMO} & DM-only & [139, -, -] & $1.39\times10^9$ & 24.92 & - & - & - & - \\
 \textcolor{Halo2}{Halo1459 DMO} & DM-only & [139, -, -] & $1.44\times10^9$ & 25.20 & - & - & - & - \\
 \textcolor{Halo5}{Halo600 DMO} & DM-only & [139, -, -] & $3.42\times10^9$ & 33.62 & - & - & - & - \\
 \textcolor{Halo4}{Halo605 DMO} & DM-only & [139, -, -] & $3.33\times10^9$ & 33.31 & - & - & - & - \\
 \textcolor{Halo3}{Halo624 DMO} & DM-only & [139, -, -] & $3.58\times10^9$ & 34.13 & - & - & - & - \\
 \hline
 \textcolor{GM}{Halo1459 DMO GM:Later} & DM-only & [139, -, -] & $1.47\times10^9$ & 25.34 & - & - & - & - \\
 \textcolor{GM2}{Halo1459 DMO GM:Latest} & DM-only & [139, -, -] & $1.47\times10^9$ & 23.34 & - & - & - & - \\
 \hline
 \textcolor{Halo1}{Halo1445} & Agertz+2020 & [117, 18, 300] & $1.32\times10^9$ & 23.10 & $1.35\times10^{5}$ & 100.80 & -6.93 \\
 \textcolor{Halo2}{Halo1459} & Agertz+2020 & [117, 18, 300] & $1.43\times10^9$ & 23.75 & $3.77\times10^{5}$ & 98.80 & -8.03 \\
 \textcolor{Halo5}{Halo600} & Agertz+2020 & [117, 18, 300] & $2.65\times10^9$ & 31.17 & $9.84\times10^{5}$ & 109.65 & -9.19 \\
 \textcolor{Halo4}{Halo605} & Agertz+2020 & [117, 18, 300] & $3.20\times10^9$ & 31.08 & $1.93\times10^6$ & 101.83 & -9.84 \\
 \textcolor{Halo3}{Halo624} & Agertz+2020 & [117, 18, 300] & $3.23\times10^9$ & 29.18 & $1.08\times10^{6}$ & 107.04 & -9.44 \\
 \hline
 \textcolor{GM}{Halo1459 GM:Later} & Agertz+2020 & [117, 18, 300] & $1.43\times10^9$ & 23.73 & $1.11\times10^{5}$ & 168.16 & -6.70 \\
 \textcolor{GM2}{Halo1459 GM:Latest} & Agertz+2020 & [117, 18, 300] & $1.38\times10^9$ & 23.45 & $8.65\times10^{4}$ & 303.80 & -6.44 \\
 \hline\hline
\end{tabular}
\label{tab:sims}
\end{table*}

In addition, to further test our numerical results, we use the {\sc griffin} code \citep{2014ComAC...1....1D} to run a controlled investigation into the effects of the merger history on the dark matter halo of one of our {\sc ramses} simulations. The motivation for these additional simulations will be made clear in Section~\ref{griffin}. {\sc Griffin} is a high performance $N$-body integrator that exploits the fast multipole method (FMM), and is an ideal tool to compare against {\sc ramses} because it is based on a fundamentally different numerical integration scheme that has comparable force accuracy to direct summation codes \citep[e.g.][]{2014ComAC...1....1D, 2017MNRAS.464.2301G, 2020MNRAS.tmp.2024N, 2020arXiv201104663N}. These simulations are run at both the equivalent and $10\times$ better mass resolution compared to our {\sc ramses} runs, with the former being used as a convergence study. We employ a force softening length of 10\,pc in all of these simulations.  Additional tests with a softening length of 5\,pc showed no measurable change in the results.

\subsection{Halo finding}

We use the {\sc hop} halo finder \citep{1998ApJ...498..137E} to identify all distinct bound structures in each simulation output. {\sc Hop} does not identify haloes within haloes (subhaloes), and so, where necessary, our analysis is supplemented with the {\sc ahf} (Amiga Halo Finder) \citep{2009ApJS..182..608K}. Merger trees and halo properties are calculated using {\sc pynbody} \citep{pynbody} and {\sc tangos} \citep{tangos}, respectively.
We locate the centre of each bound structure using the shrinking spheres method of \citet{2003MNRAS.338...14P}, performed exclusively on the dark matter component, as implemented in {\sc pynbody}. Results from our dark matter only (DMO) simulations are in all cases corrected for the universal baryon fraction. \par


Throughout this paper, the virial radius, $r_{200 \rm c}$, is defined as the spherical region that is at least 200 times the critical mass density of the universe at that redshift. The halo mass $M_{200 \rm c}$, is then the total mass of all matter contained within that radius. \par


\section{An overview of the simulations} \label{overview}

\begin{figure*}
\centering
\setlength\tabcolsep{2pt}%
\includegraphics[ trim={0cm 0cm 0cm 0cm}, clip, width=\linewidth, keepaspectratio]{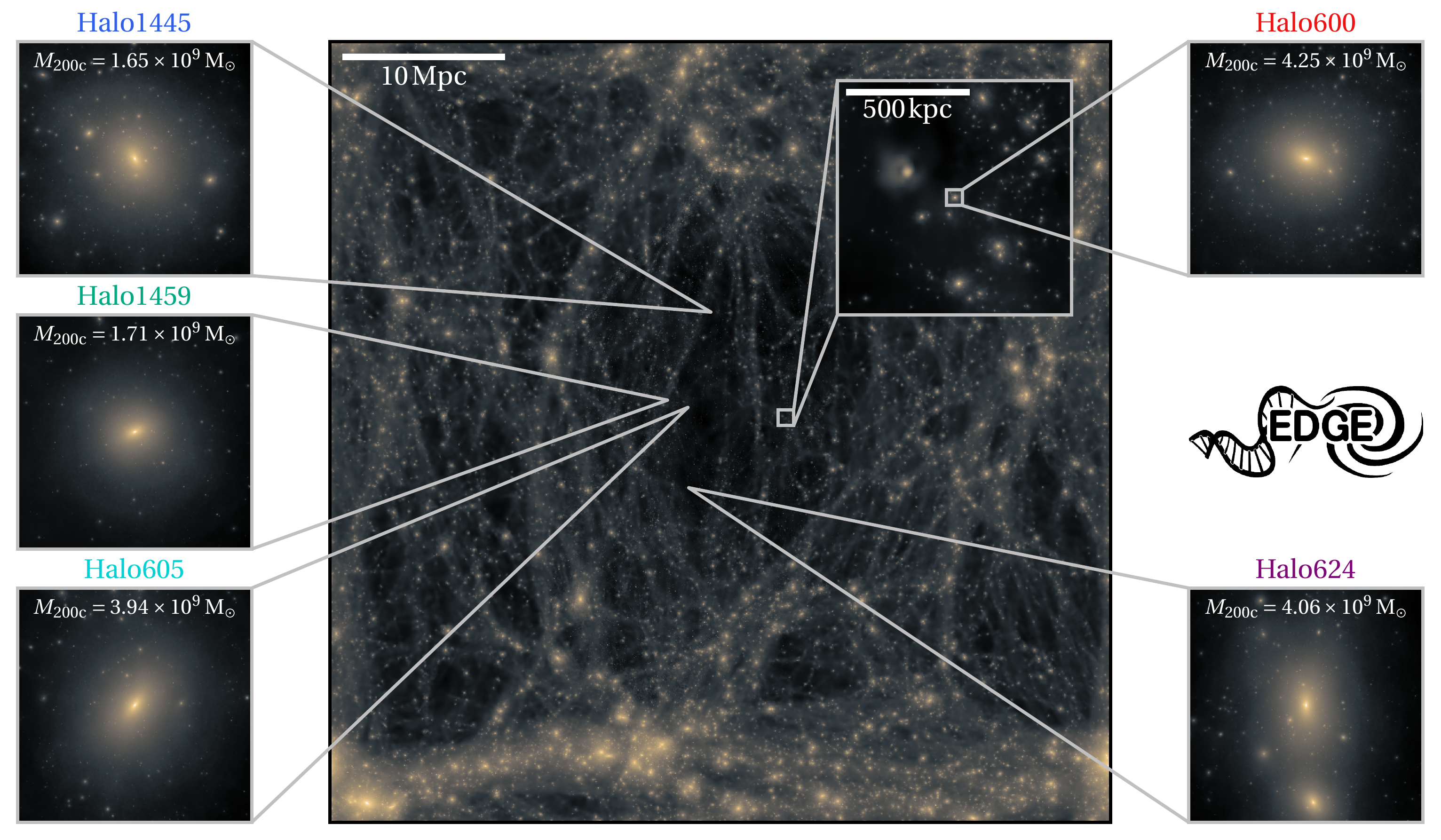}
\caption{The location of the EDGE dark matter haloes selected for higher resolution resimulation from a lower resolution void at $z=0$. Each panel shows a surface density plot of a cube. The zoomed panels for each halo are taken from the `DMO' simulations out to $r_{200 \rm c}$ (see Table \ref{tab:sims}), and show the corresponding $M_{200 \rm c}$ mass, as marked. For Halo600, there is a partial-zoom to help illustrate how small the selected haloes are in comparison to the total box size (50\,Mpc).}
\label{fig:void}
\end{figure*}

\begin{figure*}
\centering
\setlength\tabcolsep{2pt}%
\includegraphics[ trim={0cm 0cm 0cm 0cm}, clip, height=0.93\textheight, keepaspectratio]{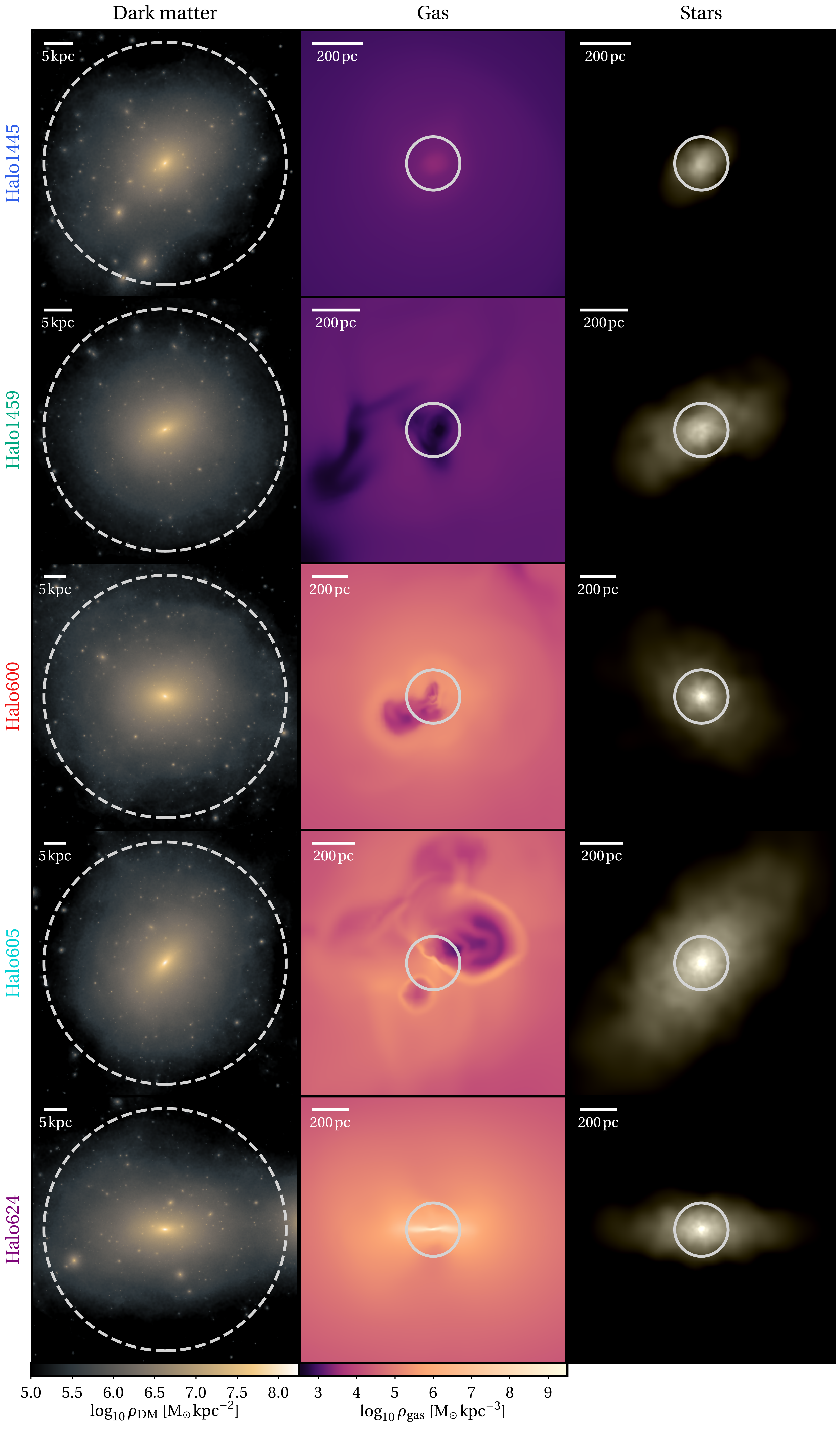}
\caption{A visual representation of the high resolution baryonic simulations in Table \ref{tab:sims}. \textit{Left panels}: The dark matter surface density out to $r_{200 \rm c}$ (dashed circle). \textit{Middle panels}: The gas density averaged along a 0.2\,kpc slice through $z$. The 3D half light radius is indicated by the solid circle. \textit{Right panels}: The halo stars are rendered in {\sc pynbody} with the $i$, $v$ and $u$-bands over the range $23 \leq \text{mag}\,\text{arcsec}^{-2} \leq 28$, shown at the same scale as the gas density panels. All images are oriented side-on along the angular momentum vector of the cool gas ($<10^4$\,K) within 1\,kpc of the halo centre. The physical size of each frame is indicated by a scale bar in the top left corner.}
\label{fig:mugshots}
\end{figure*}

\begin{figure*}
\centering
\setlength\tabcolsep{2pt}%
\includegraphics[ trim={0cm 1cm 0cm 0cm}, width=\linewidth, keepaspectratio]{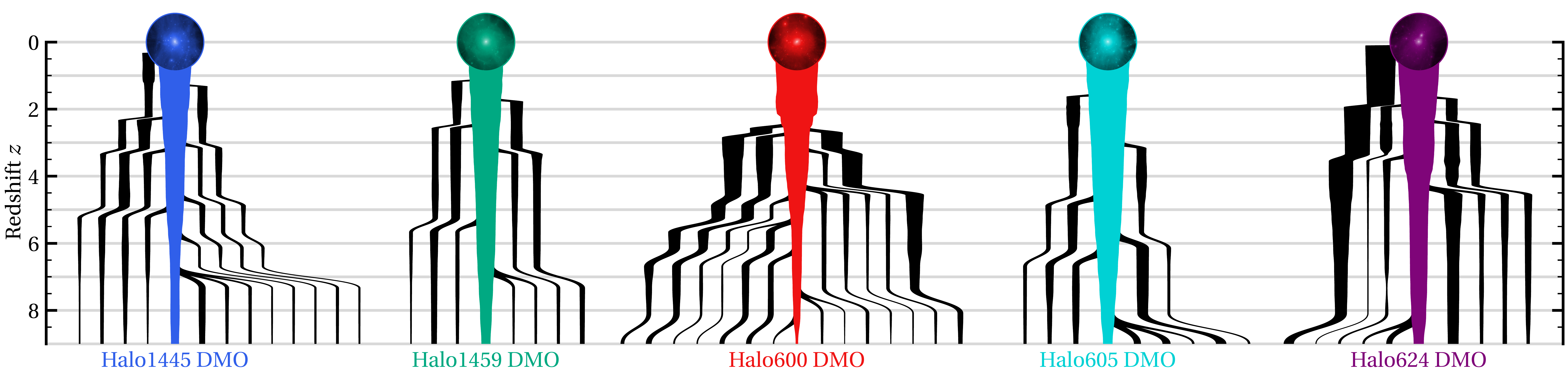}
\caption{The accretion history of each halo as taken from the `DMO' simulations, where $<1/30$ mass ratio mergers and sub-mergers are excluded for clarity. The central coloured branch is the main progenitor halo and the line thickness represents the $M_{200 \rm c}$ mass of each dark matter halo.}
\label{fig:tree}
\end{figure*}

Several dwarfs presented in this paper have been discussed already in previous EDGE collaboration papers\footnote{Halo1459, Halo1459 GM:Later and Halo1459 GM:Latest appear in \citet{2019ApJ...886L...3R}. Halo600, Halo605 and Halo624 appear in \citet{2020MNRAS.497.1508R}}, however run at a lower `fiducial' resolution ($m_{\rm DM} = 1112\,\text{M}_{\odot}$). Here, since we are interested in resolving potentially very small dark matter cores, the same dwarfs are resimulated at what we called `hires' resolution in \citet{2020MNRAS.491.1656A} ($m_{\rm DM} = 117\,\text{M}_{\odot}$). We show convergence tests between these two resolutions in Appendix \ref{appendix:c}, demonstrating that our results presented here do not depend on resolution. \par

Figure \ref{fig:void} shows a surface density plot for the dark matter of the total void region from which our initial conditions were selected. The locations for each of our haloes in Table \ref{tab:sims} are indicated with zoomed surface density plots, with images taken from the `DMO' simulations. This highlights that our haloes are chosen from a particularly under-dense region, without any major cosmic structure in the near vicinity.

Figure \ref{fig:mugshots} shows a visual representation of each high resolution baryonic simulation at $z=0$.
The left-most panels show the centred dark matter surface density out to $r_{200 \rm c}$, which is indicated by a dashed circle.
The middle panels show the central gas density averaged in a 0.2\,kpc thick slice through the $z$-axis. The plot is zoomed into the inner five half light radii, where the 3D half light radius is marked with a solid circle. Each halo is oriented side-on on the angular momentum vector of the central cold gas ($<10^4$\,K), where available, which represents the central gas disk if it is present.
The right-most panels are a {\sc pynbody} rendering of the halo stars. \par

For Halo1445 and Halo1459, the gas is extremely under-dense and shows little structure, with the exception of some mild stirring due to late Type Ia SNe (most apparent in Halo1459). Both Halo600 and Halo605 have denser gas, with bubbles forming due to ongoing bursty star formation. Halo624 is the only galaxy to form a structured gas disc, which is both dense and rotating at $z=0$. We will present a detailed study of the observational properties of these simulated dwarfs in forthcoming papers. In this paper, we focus on their dark matter content and structure. \par

In Figure \ref{fig:tree}, we present merger trees for each of our haloes using the `DMO' simulations. For simplicity, only major mergers onto the main progenitor (coloured line) are shown. The line thickness is representative of each halo mass. The general form of these merger trees are identical for different resolutions and physics, with the one exception that the final merger in Halo624 DMO occurs just after $z=0$ in Halo624. The present-day main progenitor halo is not necessarily the most massive halo at all times (for instance, Halo600 DMO at $z=6$). This is because the main progenitor is defined as the most massive halo at the time of each merger. \par

\section{Results}

\subsection{Core formation from gas flows} \label{gasflows}

\begin{figure*}
\centering
\setlength\tabcolsep{2pt}%
\includegraphics[ trim={0cm 0cm 0cm 0cm}, clip, width=\linewidth, keepaspectratio]{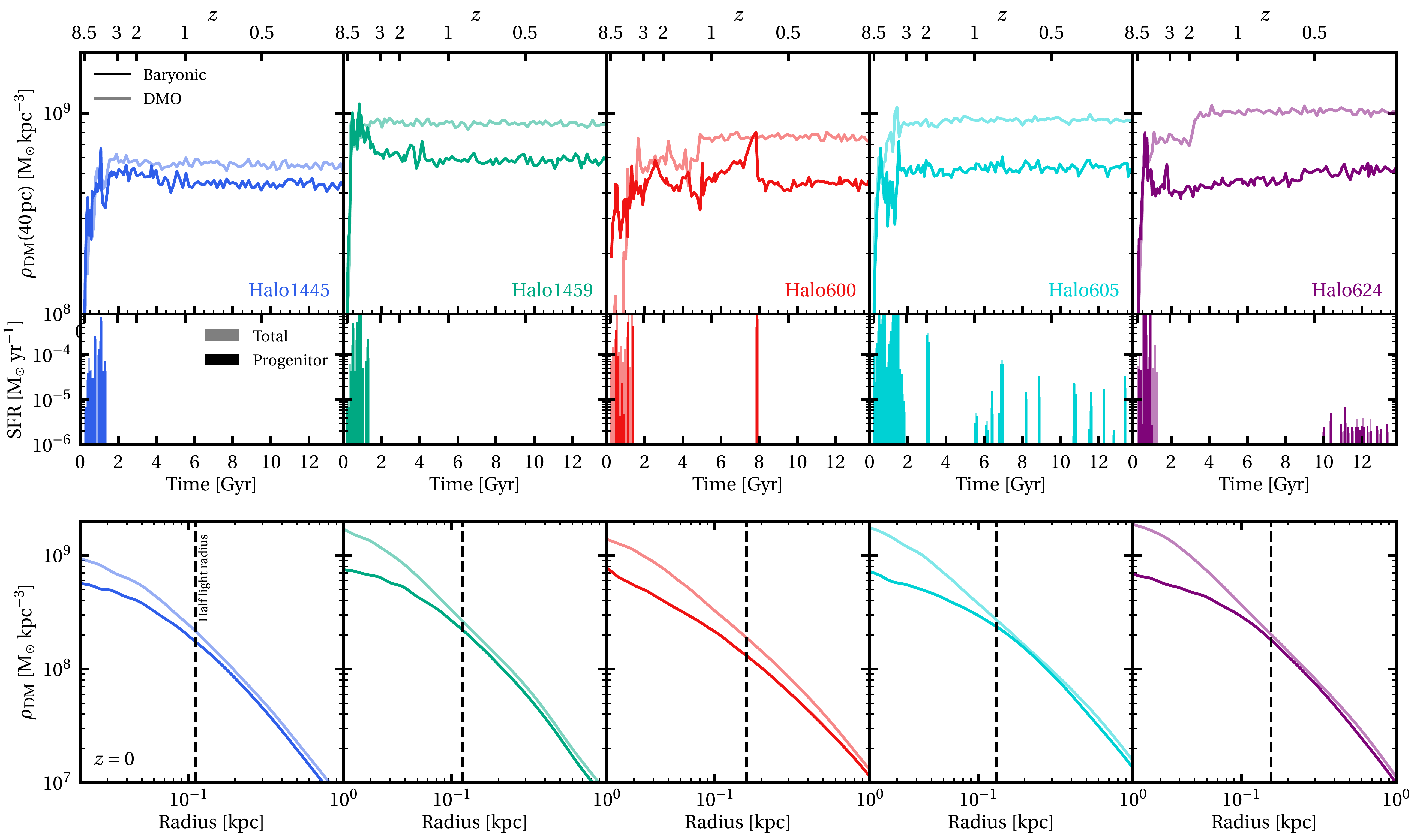}
\caption{\textit{Upper panels:} The evolution of the 3D dark matter density at $40\,$pc in the main progenitor halo for the high resolution dwarf galaxies. The opaque lines show simulations run with baryonic physics, whereas the faint lines show pure dark matter (DMO) simulations using the same initial conditions. Our results are qualitatively similar when selecting the density at different inner radii.
\textit{Middle panels:} The star formation rate of the baryonic simulations averaged over bins of $100\,$Myr. The opaque bars show stars formed within $r_{200 \rm c}$ of the main progenitor, whereas fainter bars include stars that are brought in with mergers.
\textit{Lower panels:} A comparison of the 3D dark matter density profiles between the baryonic and DMO simulations at $z=0$. The black dashed lines mark the 3D half light radii in each case.}
\label{fig:central_density_evolution}
\end{figure*}

In Figure \ref{fig:central_density_evolution}, we show the evolution of the central dark matter density for all of our EDGE simulations. In the upper panels, the inner 3D dark matter density is plotted at $40\,$pc. This approaches the inner limit of the region that we consider numerically resolved (see Appendix \ref{appendix:a}). The opaque lines are the results for the baryonic simulations, whereas faint lines are the results for the pure dark matter (DMO) simulations. The corresponding star formation rates of both the main progenitor and all progenitors for the baryonic simulation are plotted in the middle panels, averaged over 100\,Myr bins. Included in the lower panels are the 3D density profiles for both baryonic and DMO simulations at $z=0$. \par

From Figure \ref{fig:central_density_evolution}, we see that the inner dark matter density of the baryonic simulations -- that include star formation, gas cooling and stellar feedback -- is in all cases lower than the DMO simulations after $\sim1$\,Gyr. This disparity is driven by the early star formation period seen in all our dwarfs. The reduction in inner density occurs over this star forming period, as expected from the gas-flow mechanism. Star formation reignites at later times in some of our dwarfs, a result that is explored in detail for our fiducial resolution simulations in \citet{2020MNRAS.497.1508R}. The intensity of this late rejuvenation is insufficient to drive a further reduction of the inner density in Halo624 and Halo605, but is great enough in Halo600 to manifest as a sudden dip in the inner density. However, the late rejuvenation in Halo600 lacks the continuous bursts of star formation necessary to grow a large core \citep{2005MNRAS.356..107R, 2012MNRAS.421.3464P}. \par

The lower panels of Figure \ref{fig:central_density_evolution} show at which radius the density profile slopes in the baryonic simulations depart from the DMO simulations. In all cases, the bulk of the profile flattening occurs within the 3D half light radius (black dashed line) at approximately 100\,pc. This is also consistent with prior work on the gas flow mechanism (see e.g. \citealt{2015MNRAS.454.2981C,2016MNRAS.459.2573R}; and Section \ref{sec:intro}). \par 

\begin{figure}
\centering
\setlength\tabcolsep{2pt}%
\includegraphics[ trim={0cm 0cm 0cm 0cm}, clip, width=\columnwidth, keepaspectratio]{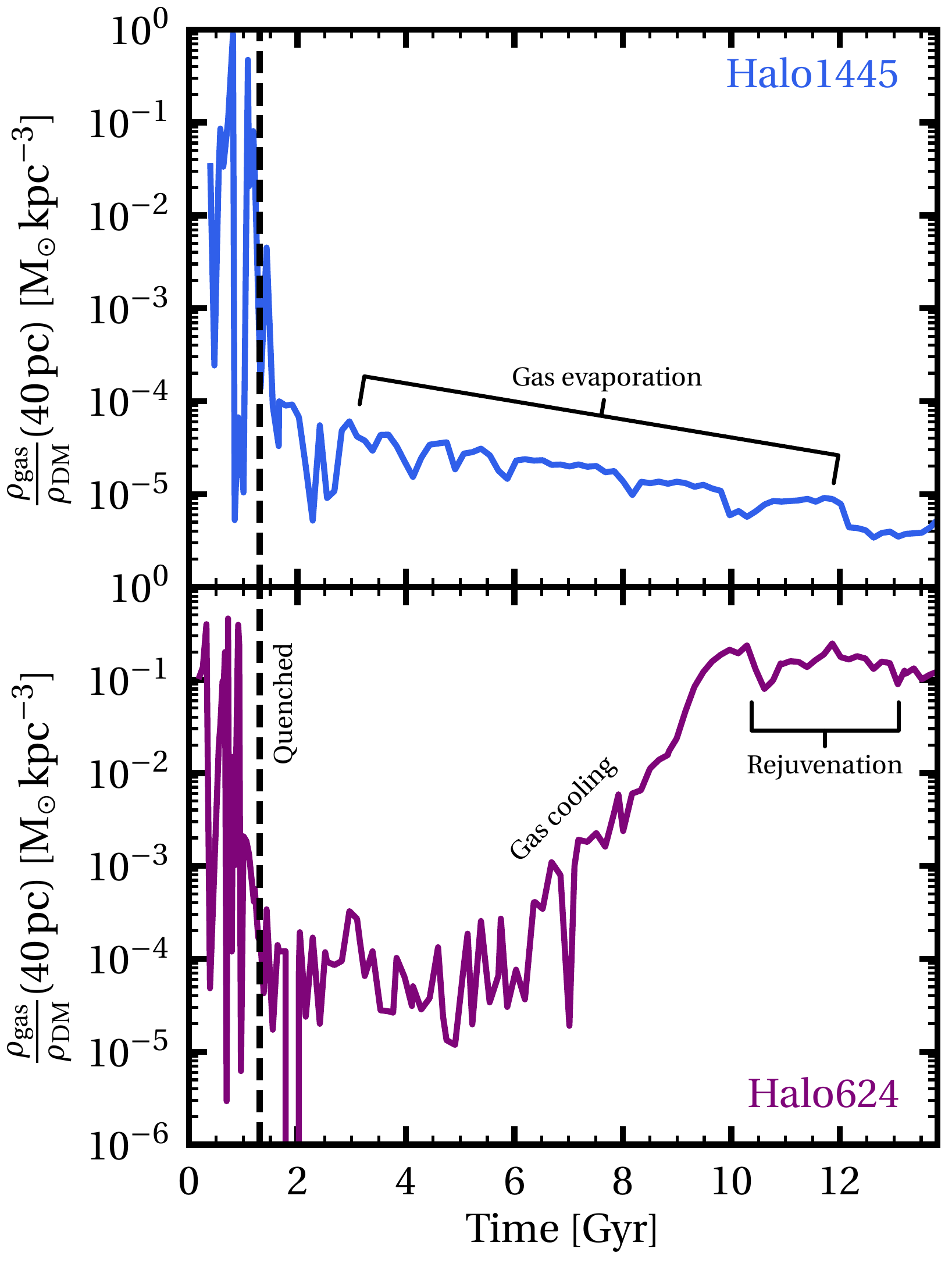}
\caption{Evolution of the inner gas-to-dark matter density ratio at 40\,pc ($\rho_{\rm gas} / \rho_{\rm DM}$) for two representative EDGE dwarfs. Prior to quenching by reionisation (black dashed line), there are large fluctuations in $\rho{\rm gas} / \rho{\rm DM}$ caused by repeated cycles of gas cooling, star formation and stellar feedback. These fluctuations excite `dark matter heating' that lowers the central dark matter density (Figure \ref{fig:central_density_evolution}). Halo1445 experiences no further star formation or dark matter heating after it quenches. By contrast, Halo624 grows in mass, accumulating cold gas and rejuvenating its star formation after $\sim 10$\,Gyr. Subsequent fluctuations are orders of magnitude smaller than at early times, and so there is no late time dark matter heating.} \label{fig:gas_ratio}
\end{figure}

In Figure \ref{fig:gas_ratio}, we show the gas-to-dark matter central density ratio ($\rho_{\rm gas} / \rho_{\rm DM}$) for two representative dwarf galaxies. This is computed at time intervals of $\Delta a = 0.01$ (the cadence of our simulation outputs), and so the true peak $\rho_{\rm gas} / \rho_{\rm DM}$ may be higher. The upper panel shows Halo1445, which is quenched permanently by $z=4$ due to the effects of reionisation. There are large-scale fluctuations in the density ratio before quenching, corresponding to gas outflow/inflow triggered by bursty star formation. These fluctuations cease after quenching, and the gas gradually photo-evaporates. The lower panel shows Halo624, which is a more massive dwarf that is able to rejuvenate at late times. As in Halo1445, there are large-scale density ratio fluctuations at early times that diminish after the initial quenching. However, as the halo grows more massive it is able to increase its central gas density \citep[see][]{2020MNRAS.497.1508R}. Star formation then rejuvenates after 10\,Gyr, but the intensity of this late star formation is not sufficient to drive large enough fluctuations in $\rho_{\rm gas} / \rho_{\rm DM}$ to further heat the central dark matter.\par

\subsection{Core formation from minor mergers} \label{mergers}

\begin{figure}
\centering
\setlength\tabcolsep{2pt}%
\includegraphics[ trim={0cm 0cm 0cm 0cm}, clip, width=\columnwidth, keepaspectratio]{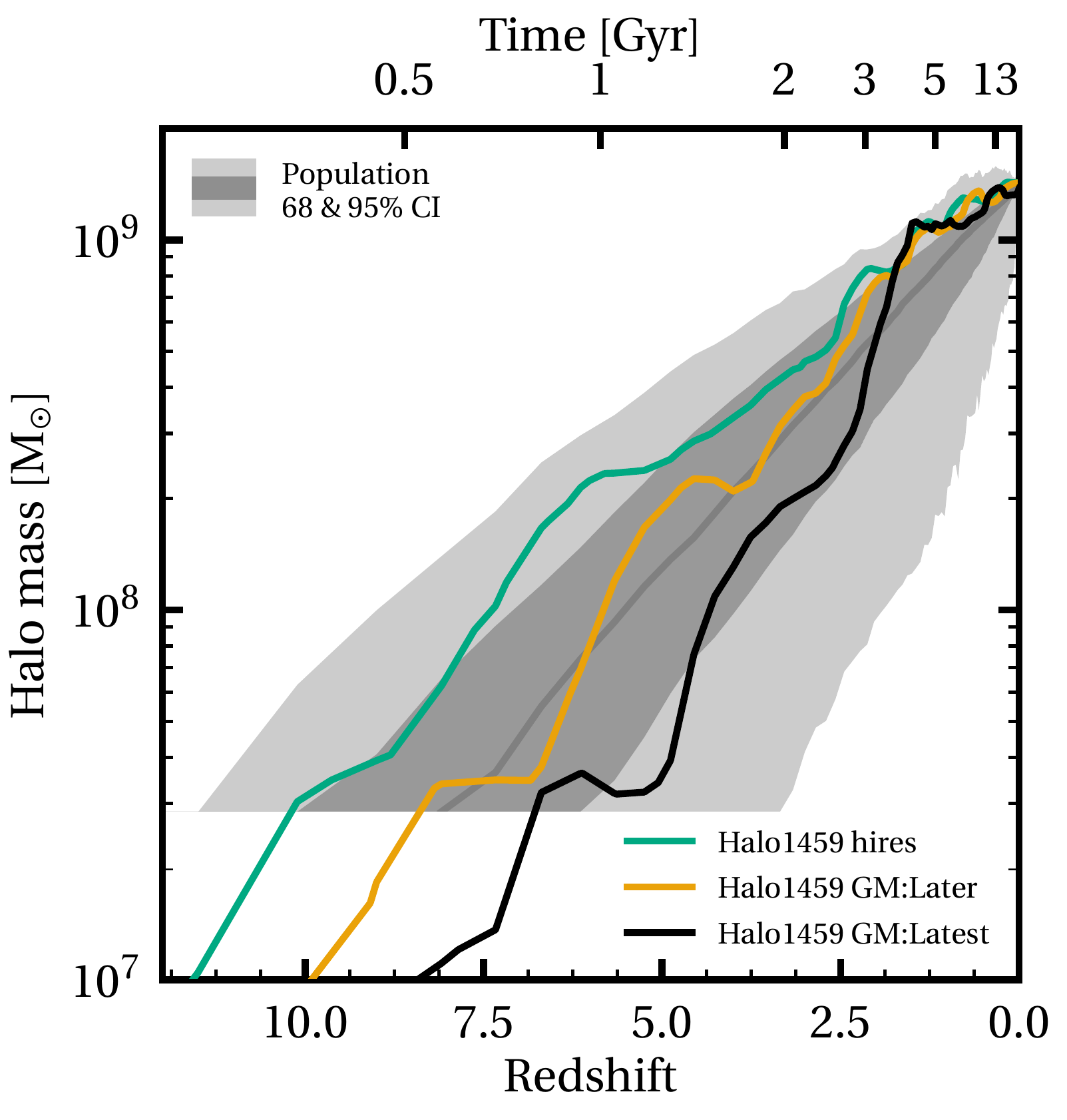}
\caption{The halo mass growth history of Halo1459 and two modified variants which we call Halo1459 GM:Later and Halo1459 GM:Latest. Included are grey bands indicating the 68 and 95 per cent scatter for the mass growth histories of haloes throughout our lower resolution void simulation. The bands are truncated at $\sim 3 \times 10^7$ due to the resolution limit of the void simulation. This shows that our modified haloes are within the expected scatter of assembly histories in $\Lambda$CDM.}
\label{fig:gm_mass_growth}
\end{figure}

\begin{figure}
\centering
\setlength\tabcolsep{2pt}%
\includegraphics[ trim={0cm 0.5cm 0cm 0cm}, clip=False, width=\columnwidth, keepaspectratio]{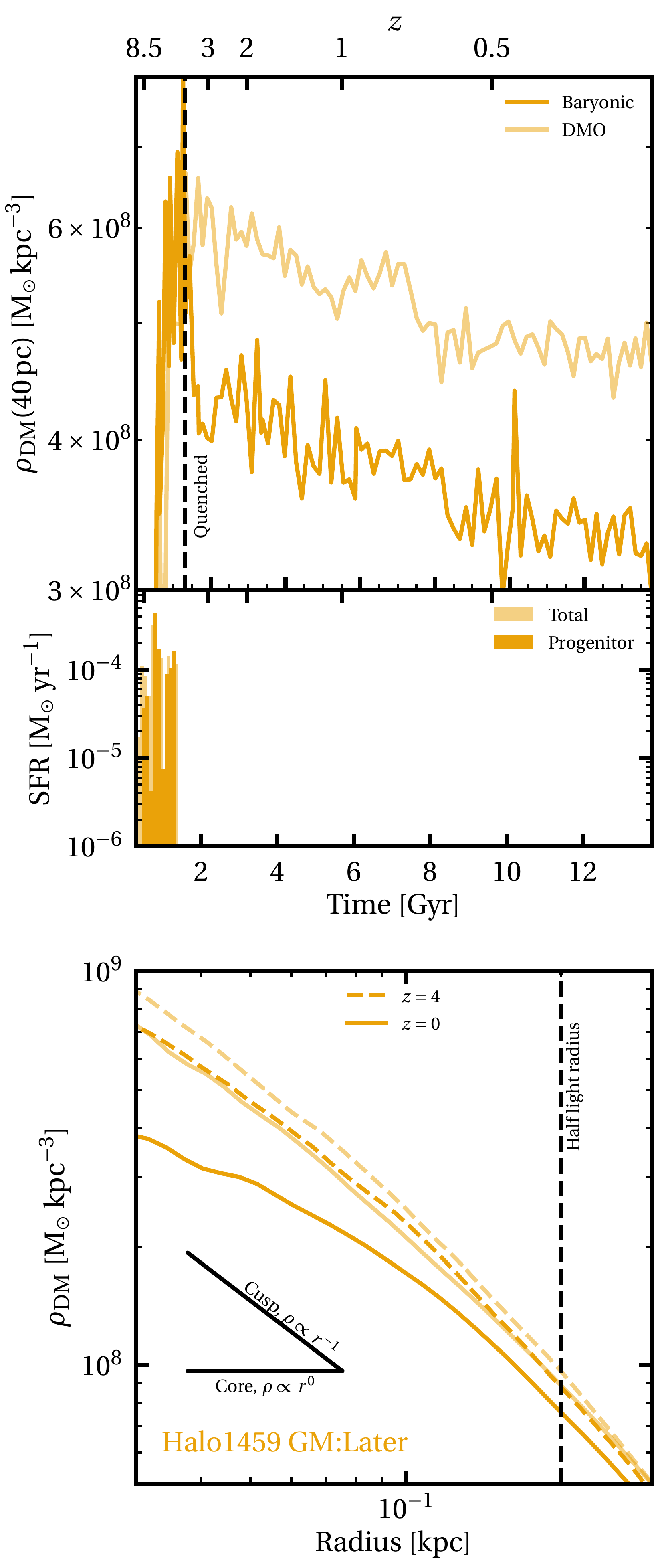}
\caption{\textit{Upper panel:} The evolution of the 3D dark matter density at 40\,pc, but for a modified simulation that has a delayed formation history (Halo1459 GM:Later). A black dashed line marks the approximate time at which the galaxy is permanently quenched.
\textit{Middle panel:} The star formation rate of the baryonic simulation, averaged over bins of $100\,$Myr. Opaque bars are stars formed within $r_{200 \rm c}$ of the main progenitor, whereas faint bars include stars that are brought in with mergers.
\textit{Lower panel:} A comparison of the 3D dark matter density profiles between the baryonic and DMO simulations at both $z=4$ (after quenching) and $z=0$. A black dashed line marks the 3D half light radius at $z=0$.}
\label{fig:coreformation_mergers}
\end{figure}

Our results so far are in line with previous studies in the literature that suggest that forming kpc scale dark matter cores becomes inefficient in ultra-faint dwarfs \citep{2012ApJ...759L..42P, 2013MNRAS.433.3539G, 2014MNRAS.441.2986D, 2014ApJ...789L..17M, 2015ApJ...806..229M, 2016MNRAS.456.3542T, 2016MNRAS.459.2573R}. However, the puzzle of apparent dark matter cores in at least some ultra-faints remains \citep{2017ApJ...844...64A, 2018MNRAS.476.3124C}, Section~\ref{sec:intro}. \par

Perhaps the most compelling case for a small dark core in an ultra-faint dwarf comes from the survival of a lone and extended star cluster, offset from the centre of the Eridanus II galaxy \citep{2017ApJ...844...64A,2018MNRAS.476.3124C,2020arXiv201200043S}. Eridanus II is substantially more extended than any of our reference EDGE dwarfs, with $R_{\rm half} = 299 \pm 12$\,pc. Using the genetic modification approach \citep{2016MNRAS.455..974R, 2018MNRAS.474...45R, 2020arXiv200601841S}, we can create alternative mass accretion histories for a galaxy. \citet{2019ApJ...886L...3R} modified one of the fiducial resolution EDGE dwarfs, Halo1459, such that it assembled later. They found that this leads to a larger, lower surface brightness, and lower metallicity dwarf -- more similar to Eridanus II. As such, in this section, we resimulate the genetically modified later forming dwarfs from \citet{2019ApJ...886L...3R} at higher resolution to study how late formation impacts the central dark matter density.

We run two modifications of Halo1459 which we call Halo1459 GM:Later and Halo1459GM:Latest (Table \ref{tab:sims}). The assembly histories of these modifications were already shown at our fiducial resolution in Figure 1 of \citet{2019ApJ...886L...3R}; they are indistinguishable from the higher resolution trajectories which we show in Figure \ref{fig:gm_mass_growth}. The modified haloes are approximately $3\times$ ($2\times$) less massive at $z=8.5$ for Halo1459 GM:Later (Halo1459 GM:Latest), but grow to the same mass within a 4 per cent margin by $z=0$.\par

In the upper panel of Figure \ref{fig:coreformation_mergers}, we show the evolution of the central density in Halo1459 GM:Later. Now, the inner dark matter density continues to fall long after star formation has ceased. This occurs in both the simulation with baryonic physics and in the DMO simulation. The baryonic simulation is also extremely gas deficient after quenching, with $M_{\rm gas} / M_{\rm DM}(<r_{\rm half light}) < 10^{-5}$, so any gas flows driven by residual feedback from old stars \citep[as in][]{2020MNRAS.497.1508R} have a negligible impact on the overall mass distribution. This reduction in the central density is not seen to the same extent in Halo1459 GM:Latest. We will discuss the reasons for this in Section~\ref{core destruction}.

The lower panel of Figure \ref{fig:coreformation_mergers} shows the radial dark matter density profiles of Halo1459 GM:Later (darker) and Halo1459 DMO GM:Later (lighter) at $z=4$ (dashed) and $z=0$ (solid). A schematic shows the expected density slopes for a cusp and a core, indicating that whilst the central density slope has declined by $z=0$, it still has a slope of $\rho \propto r^{-0.5}$. \par

It is important to rule out the possibility that any apparent dark matter heating is caused by numerical relaxation. In Appendix \ref{appendix:a}, we calculate the `relaxation radius', $r_{\rm relax}$, for our EDGE simulations. This is the radius inside which the numerical relaxation time is equal to the simulation run time and so numerical relaxation will become important. From this calculation, we conclude that our simulations should still be well-resolved above $r_{\rm relax} > 25\,$pc at $z=0$, yet the density clearly evolves on scales larger than this at \textit{all} times. Therefore, there must be some other mechanism by which the central dark matter density is lowering. It is well established that dynamical heating from dense clumps can contribute to core formation (e.g. \citealt{2001ApJ...560..636E, 2004MNRAS.353..829M, 2009ApJ...702.1250R, 2010ApJ...725.1707G, 2015MNRAS.446.1820N}; and Section \ref{sec:intro}). In Halo1495 GM:Later, the only dense clumps available to drive such a process at late times are merging dark matter subhaloes.\par

To investigate the veracity of the above late time dark matter heating, and to explore whether merging dark matter subhaloes are indeed the culprit, in the next section we resimulate the sequence of mergers that form Halo1459 GM:Later using the non-cosmological $N$-body FMM code {\sc griffin} \citep{2014ComAC...1....1D}, as described in Section~\ref{simulation methods}. This allows us to isolate the heating effects of minor mergers in a non-cosmological setting, and to verify that numerical effects unique to {\sc ramses} are not responsible for the heating. \par

\subsubsection{Testing minor merger induced dark matter core formation with the {\sc griffin} code} \label{griffin}

In this section, we use the {\sc griffin} $N$-body code to reproduce the assembly history of Halo1459 GM:Later in a controlled manner. For this, we simulate a series of halo mergers based upon profile fits to the reference {\sc ramses} simulation. We use the coreNFW profile as in \citet{2016MNRAS.459.2573R}, which is a NFW profile \citep{navarro1997universal} adapted to include a parameter $n$ that controls the flatness of the central density slope.

We first fit the spherically symmetric dark matter density profile of the main progenitor halo at $z=4$. By this time, the central dark matter halo is well-established and star formation has permanently quenched. The total stellar mass is low ($\sim10^5\,\text{M}_{\odot}$) and the gas content is negligible, so the system can be safely resimulated using exclusively its dark matter component. \par

Mergers are defined based on every halo identified with {\sc hop} at $z=4$, on the condition that they contain at least 800 dark matter particles and are destined to merge with the main progenitor. Each merging halo is tracked until the snapshot prior to infall (defined as the point where the merger crosses over $r_{200 \rm c}$ of the main progenitor), by which time many of them have coalesced. Despite the large number of individual haloes at $z=4$, there are a manageable 38 distinct subhaloes at the time of merging. \par

For each of these mergers, we perform a spherically symmetric coreNFW profile fit and generate initial conditions using {\sc agama} \citep{2018ascl.soft05008V}. We use a multipole potential approximation with one hundred grid nodes, an isotropic velocity distribution function, and model the density out to $3 \times r_{200 \rm c}$ for each coreNFW profile fit (which is necessary to ensure there is sufficient dynamical friction between interacting haloes at large radii;  \citealt[e.g.][]{2008MNRAS.389.1041R}). A further improvement could include fitting the halo triaxiality, but should not be necessarily for resolving the leading order effects of halo mergers. \par

The initial conditions of the main progenitor are integrated forward in time with the {\sc griffin} code, as described in Section~\ref{simulation methods}. As the simulation reaches the time of each merger infall, the corresponding initial conditions are inserted at the same phase-space location as in the reference {\sc ramses} simulation. In this way, the merger history of the {\sc ramses} simulation from $4\leq z\leq 0$ is recreated. \par

The above method has a few caveats. Firstly, the number density of mergers in a $\Lambda$CDM cosmology increases with decreasing halo mass \citep{2009ApJ...702.1005S, 2015MNRAS.449...49R}. Therefore, a significant amount of mass accretion is neglected by only considering mergers above a certain mass threshold. This yields a final $z=0$ mass roughly $1.5\times$ less that of the original {\sc ramses} simulation, although much of this missing mass is located in the halo outskirts ($\sim80$ per cent of the missing mass is exterior to 5\,kpc). We also see that the orbits of merging haloes begin to diverge from the reference simulation after two pericentre passages, and it is already established that reproducing exact orbital behaviour of mergers is challenging \citep[e.g.][]{2010MNRAS.406.2312L}. Lastly, any mergers already within the virial radius of the main progenitor by $z=4$ are ignored, meaning that the onset of dynamical heating due to mergers may be delayed as compared to the reference simulation.\par

A control simulation without mergers was also run to distinguish any physical reduction in density from numerical relaxation. Our {\sc griffin} simulations are summarised in Table \ref{tab:griffinsims}.

\begin{table}
\centering
\caption{The {\sc griffin} simulations used to investigate the late-time density reduction in Halo1459 GM:Later. From left to right, the columns give: the simulation names, whether mergers were included or the main progenitor was isolated, the dark matter particle mass resolution, and the force softening length.}
\begin{tabular}{lccccc}
 \hline
  \textbf{Name} & \textbf{Mergers} & \textbf{Resolution} & \textbf{Force softening [pc]} \\
  & & \textbf{[$m_{\rm DM}$/M$_{\odot}$]} & \\
  \hline
  Isolated low & \ding{55} & 117 & 10 \\
  Mergers low & \ding{51} & 117 & 10 \\
  Isolated & \ding{55} & 11.7 & 10 \\
  Mergers & \ding{51} & 11.7 & 10 \\
 \hline
\end{tabular}
\label{tab:griffinsims}
\end{table}

\begin{figure}
\centering
\setlength\tabcolsep{2pt}%
\includegraphics[ trim={0cm 0cm 0cm 0cm}, clip, width=\columnwidth, keepaspectratio]{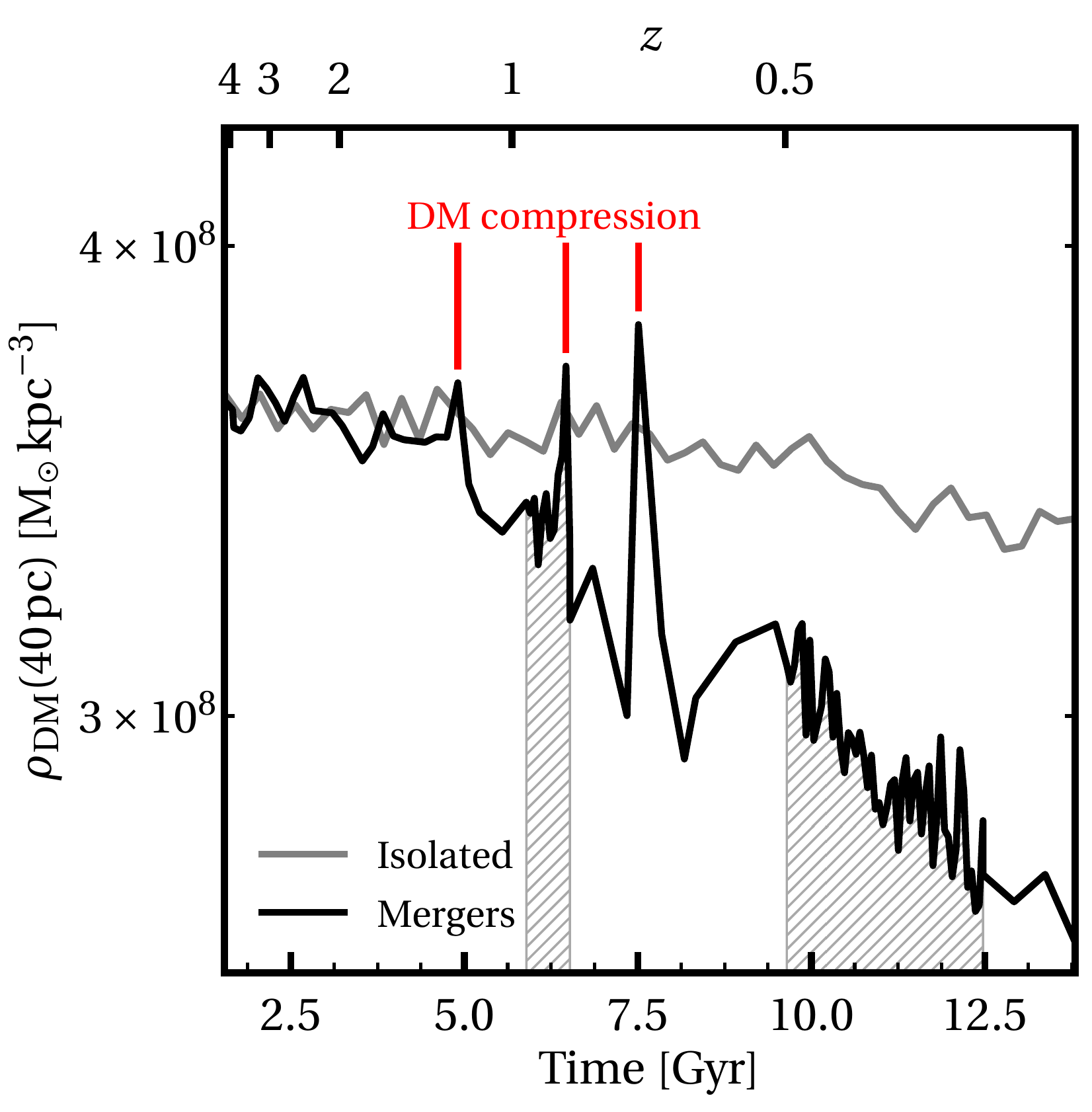}
\caption{The evolution of the 3D dark matter density at 40\,pc in the {\sc griffin} simulations. We start the $x$-axis at redshift $z=4$ to mimic the age of the reference {\sc ramses} simulation. The grey line shows a control {\sc griffin} simulation run to $z=0$ without any mergers. The black line shows the same simulation including 38 mergers. The hatched grey areas mark time intervals over which the output cadence is increased by an order of magnitude to resolve high-frequency effects. Notice that the dark matter density is lowered substantially in the simulation with mergers as compared to the control simulation. Notice, further, the three dark matter density spikes marked in red. These correspond to close subhalo passages, suggesting that the `dark matter heating' effect seen here is driven by repeated gravitational shocks from merging subhaloes (see Figure \ref{fig:griffin_merger_orbit}).}
\label{fig:griffin_density_evolution}
\end{figure}

The evolution of the central density for our isolated control (grey) and full assembly (black) {\sc griffin} simulations is shown in Figure \ref{fig:griffin_density_evolution}. At 40\,pc, there is still a small numerical heating effect in the isolated simulation (grey line). The inner density falls by $\approx 2.6\times10^7\,\text{M}_{\odot}\,\text{kpc}^{-3}$ over 12\,Gyr. However, this contrasts with a much more substantial drop in the central density of the simulation with mergers of $\approx 9.1\times10^7\,\text{M}_{\odot}\,\text{kpc}^{-3}$. The small heating present in the isolated simulation is not immediately obvious in any of our non-GM {\sc ramses} simulations, and the differences in numerical setup and lack of cosmological growth may be contributing to this. We do not run any {\sc ramses} simulations without cosmological accretion, so a strict comparison is difficult. \par

Finally, notice the three prominent `spikes' in the inner dark matter density at $\sim4.9$, $\sim 6.5$ and $\sim 7.5$\,Gyr, as marked on Figure \ref{fig:griffin_density_evolution} in red. These times correspond to close subhalo passages, suggesting that the `dark matter heating' effect is driven by repeated gravitational shocks from merging subhaloes. We discuss this further in Section \ref{discussion}.

\subsection{Cusp replenishment} \label{core destruction}

\begin{figure*}
\centering
\setlength\tabcolsep{2pt}%
\includegraphics[ trim={0cm 0cm 0cm 0cm}, clip, width=\linewidth, keepaspectratio]{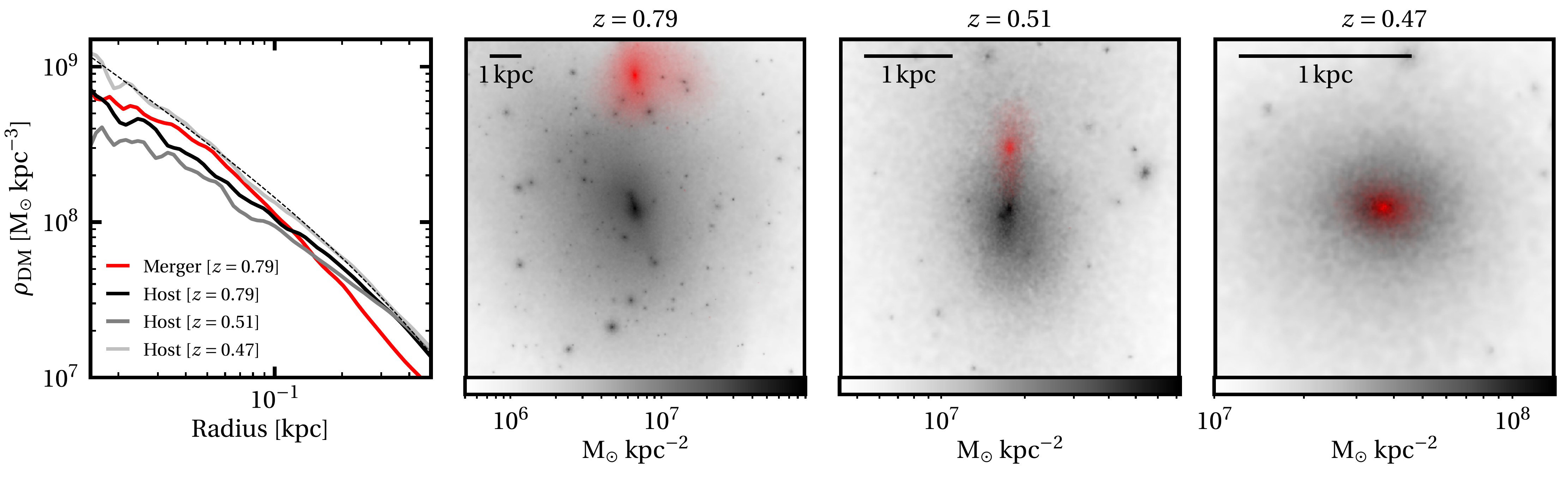}
\caption{This plot illustrates how the reintroduction of denser dark matter material to the Halo1459 GM:Latest simulation erases the effect of earlier dark matter heating. \textit{Leftmost panel:} The central 3D dark matter density profiles of a cuspy subhalo (red line) and of the parent halo evolution (black, dark grey and light grey lines). The density profile of the host at $z=0.51$ is centred on only the host particles to avoid any bias due to the merger. Removing the merger particles from the density profile calculation does not qualitatively change the results. A black dashed line represents a NFW profile fit to the parent halo at $z=0.79$ \citep{navarro1997universal}. \textit{Right panels:} The evolution of the merger system shown in three panels, where the parent halo is in grey-scale and the merging halo is red. The merging halo has permanently fallen into the central 100\,pc of the parent halo by $z=0.47$. In all cases, the merger system has been oriented such that the centres of both haloes are in the $xy$-plane.} \label{fig:cusp_reintroduction}
\end{figure*}

Along with mechanisms that flatten the central dark matter density, there are mechanisms that can rebuild it. \citet{2015MNRAS.449L..90L} investigate a scenario where dense mergers can reintroduce dynamically cold dark matter into a cored parent halo, thereby rebuilding the central density cusp. These events require that the merger is able to fall into the centre of the parent halo intact, which would demand the merging structures are resistant to the tidal disruption of the parent galaxy. \par

In Figure \ref{fig:cusp_reintroduction}, we show an example of such a merger in Halo1459 GM:Latest. The left panel shows the central dark matter densities of the merging system at several key times. The merging halo is shown in red at $z=0.79$, by which time it has permanently crossed over the $r_{\rm 200c}$ radius of the parent halo.
The parent halo is shown at the same time in black, and has already begun to depart from a primordial density cusp (black dotted line), primarily due to the action of minor mergers as in Section~\ref{mergers}.
By $z=0.51$, the dark grey line shows that the central density of the parent halo has continued to decline to its lowest point.
Finally, the light grey line shows the parent and merging halo combined at $z=0.47$, with the central density returning to a steep primordial cusp (which is coincidentally well described by the cuspy profile fit made at $z=0.79$).

The right three panels show the progression of the merger event in surface density plots, with the accreted material highlighted in red. The outer regions of the merging halo have been stripped away between $z=0.79$ and $z=0.51$, but the central density is retained. The final panel conveys how the increase in central density at $z=0.47$ is correlated with the central deposition of the merger material. \par

\section{Discussion} \label{discussion}

\subsection{Two mechanisms for fluctuating the gravitational potential}

\begin{figure}
\centering
\setlength\tabcolsep{2pt}%
\includegraphics[ trim={0cm 0cm 0cm 0cm}, clip, width=\columnwidth, keepaspectratio]{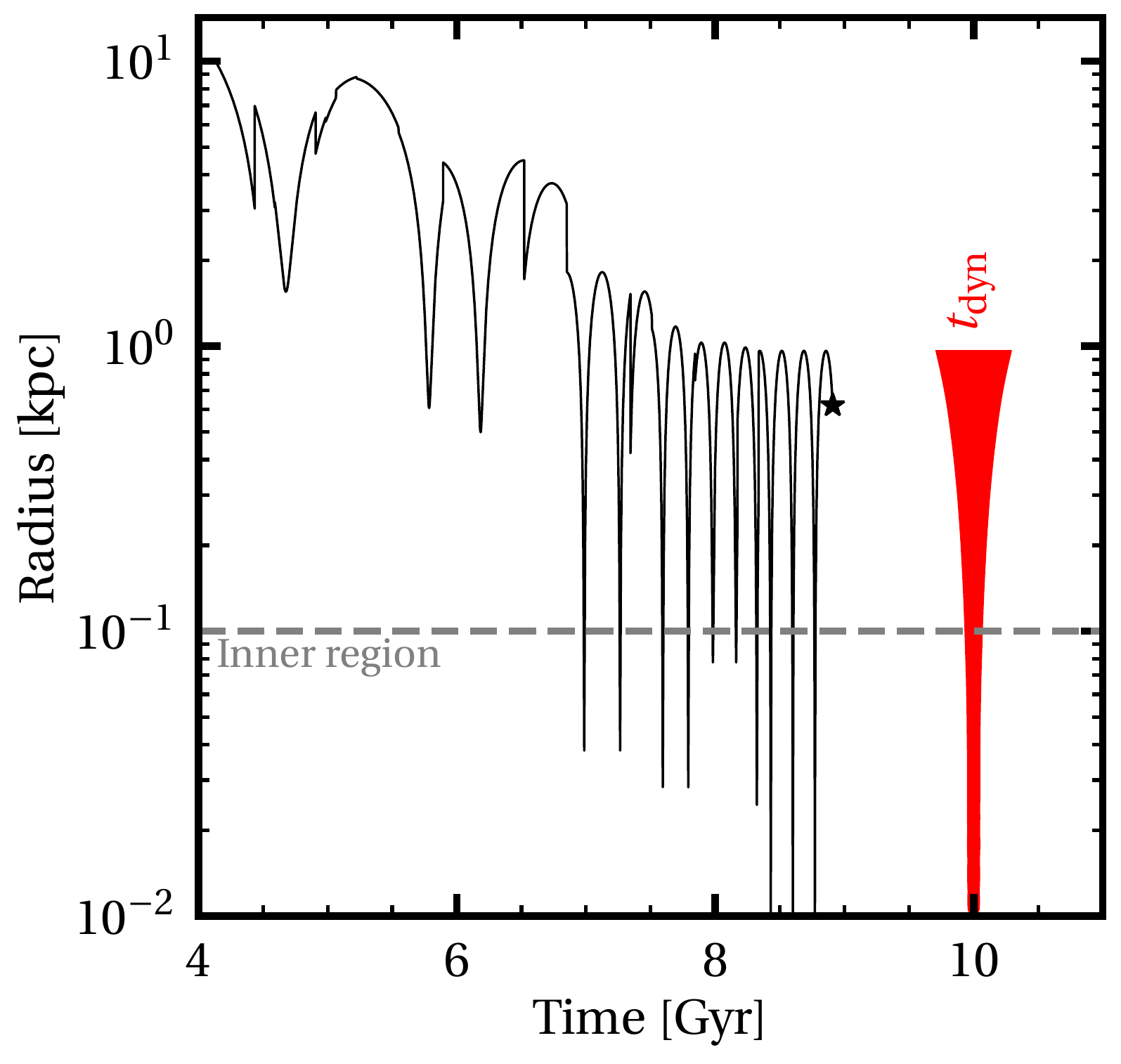}
\caption{The orbit of one example merger from our `Mergers' {\sc griffin} simulation. The orbit between each simulation output has been reconstructed with a two-body integration in {\sc agama}, assuming a spherically symmetric background potential and using a multipole fit to each simulation snapshot. Whilst these orbit reconstructions are imperfect due to perturbations from other subhaloes and triaxiality, they provide a reasonable estimate for our purposes. A red bar shows the width of one dynamical time $t_{\rm dyn} = 2 \pi \sqrt{3 / 4 \pi G \rho(r)}$ over a range of orbital radii, and a black star marks the time at which the merger dissolves. The approximate inner region is indicated at 0.1\,kpc with a horizontal dashed line.}
\label{fig:griffin_merger_orbit}
\end{figure}

We have established that gas flows driven by star formation are able to drive sufficiently large potential fluctuations to erode central density cusps in our EDGE simulations (Section \ref{gasflows}). However, this is seen only at early times when star formation rates exceed $1 \times 10^{-4}\, \text{M}_{\odot} \text{yr}^{-1}$ and fluctuate on a timescale of order the local dynamical time. Despite the late-time rejuvenation of star formation in several of our simulations, this second phase of star formation is not sufficient to drive significant gas flows and the central dark matter density is unaffected. \par

However, bursty star formation is not the only means by which the central gravitational force can be varied. We have also shown that passing subhaloes act to fluctuate the central density driving dark matter heating (Section \ref{mergers}). In Figure \ref{fig:griffin_merger_orbit}, we show an example orbit of one merging subhalo taken from our {\sc griffin} simulation. Notice that this merger repeatedly punctures the inner region of its host galaxy on a timescale shorter than the local dynamical time (red). The first close passage of this subhalo corresponds to the dark matter density spike at 6.5\,Gyr shown in Figure \ref{fig:griffin_density_evolution}. The other density spikes in that Figure correspond to close passages from different merging subhaloes. Taken together, this indicates that the late time dark matter heating is being driven by tidal shocks from the merging subhaloes on their host.

The above late time dark matter heating due to minor mergers occurs in both our baryonic and DMO {\sc ramses} simulations (Figure \ref{fig:coreformation_mergers}). However, in the DMO simulations the inner dark matter density, while lower, remains cuspy (Figure \ref{fig:coreformation_mergers}, bottom panel). By contrast, in the baryonic simulation -- and in our {\sc griffin} replica of this simulation -- these minor mergers flatten the cusp. This occurs because in these simulations, the cusp is already weakened at early times by dark matter heating due to star formation (Figure \ref{fig:coreformation_mergers}, upper panel).

Note that this minor merger induced heating has been discussed previously in the literature albeit in different contexts. \citet{2009ApJ...699L.178N} propose a mechanism by which the central concentrations of massive elliptical galaxies are reduced through repeated minor mergers, with similar effects also seen in \citet{2013MNRAS.431..767B}. And, \citet{2020MNRAS.493..320L} propose that mergers could expand the orbits of globular clusters in the Fornax dwarf spheroidal galaxy, solving a long-standing puzzle as to why they have not sunk to the centre of Fornax via dynamical friction. This same mechanism would also expand the orbits of the dark matter particles too.

\section{Conclusions} \label{conclusion}

We have presented a suite of cosmological zoom simulations of the ultra-faint dwarf galaxies performed with the adaptive mesh refinement code {\sc ramses} as part of the EDGE project. These simulations have a spatial and mass resolution of 3\,pc and $120\,\text{M}_{\odot}$, respectively, sufficient to resolve the formation of very small dark matter cores.\par

Our key result is that we uncover two distinct pathways to dark matter core formation at sub-kpc scales in the $10^9<M_{200 \rm c}/\text{M}_{\odot}<5\times10^{9}$ halo mass regime. These are able to drive reductions in the central (40\,pc) dark matter density of up to approximately a factor of two as compared to pure dark matter simulations. The first pathway is stellar feedback, in agreement with previous literature. This requires a sufficiently high star-formation rate over an extended period of time, which in our EDGE simulations only occurs at high redshift prior to reionisation. At these early times, we found that the star formation rate fluctuated on the order of the local dynamical time with an average amplitude of $\sim 1\times10^{-4}\,\text{M}_{\odot}\,\text{yr}^{-1}$. This caused the orbits of dark matter particles to migrate outwards, lowering the dwarf's inner dark matter density.\par

However, even after quenching by reionisation, we found that a second mechanism can cause dark matter cores to continue to grow: impulsive heating from minor mergers. To demonstrate this, we `genetically modified' the initial conditions for one dwarf such that it assembled later from many minor mergers. We found that, in this case, the dwarf's inner dark matter density continued to drop long after star formation ceased. We tested the veracity of this result by running an independent `replica' simulation using the {\sc griffin} $N$-body code, finding excellent agreement between the {\sc griffin} and {\sc ramses} calculations.\par

While all of our dwarfs experienced some dark matter heating prior to reionisation, we showed that dense major mergers can replenish kinematically cold dark matter, thereby reintroducing a density cusp at late times. This demonstrates that the central density of the smallest dwarf galaxies at $z=0$ is sensitive to both their mass assembly histories and their star-formation histories. This will drive stochasticity in the central dark matter densities of isolated ultra-faint dwarfs. We will study this in more detail in future work. 

Finally, none of our simulated EDGE dwarfs experienced sufficient dark matter heating to produce a flat density core. In a companion paper, we will consider whether the dark matter heating we do find in EDGE is sufficient to explain the survival and properties of the lone star cluster in Eridanus II.\par

\section*{Acknowledgements}

The author acknowledges the UKRI Science and Technology Facilities Council (STFC) for support (grant ST/R505134/1).
This project has received funding from the European Union's Horizon 2020 research and innovation programme under grant agreement No. 818085 GMGalaxies. AP was further supported by the Royal Society.
OA and MR acknowledge support from the Knut and Alice Wallenberg Foundation and the Swedish Research Council (grants 2014-5791 and 2019-04659).
MD acknowledges support by ERC-Syg 810218 WHOLE SUN.
%
This work was performed using the DiRAC Data Intensive service at Leicester, operated by the University of Leicester IT Services, which forms part of the STFC DiRAC HPC Facility (www.dirac.ac.uk). The equipment was funded by BEIS capital funding via STFC capital grants ST/K000373/1 and ST/R002363/1 and STFC DiRAC Operations grant ST/R001014/1. DiRAC is part of the National e-Infrastructure.
%

\section*{Data availability}
Data available upon request.



\bibliographystyle{mnras}
\bibliography{main} 



\appendix
\section{Deriving the relaxation radius for our EDGE simulations} \label{appendix:a}

\begin{figure*}
\centering
\setlength\tabcolsep{2pt}%
\includegraphics[ trim={0cm 0cm 0cm 0cm}, clip, width=\linewidth, keepaspectratio]{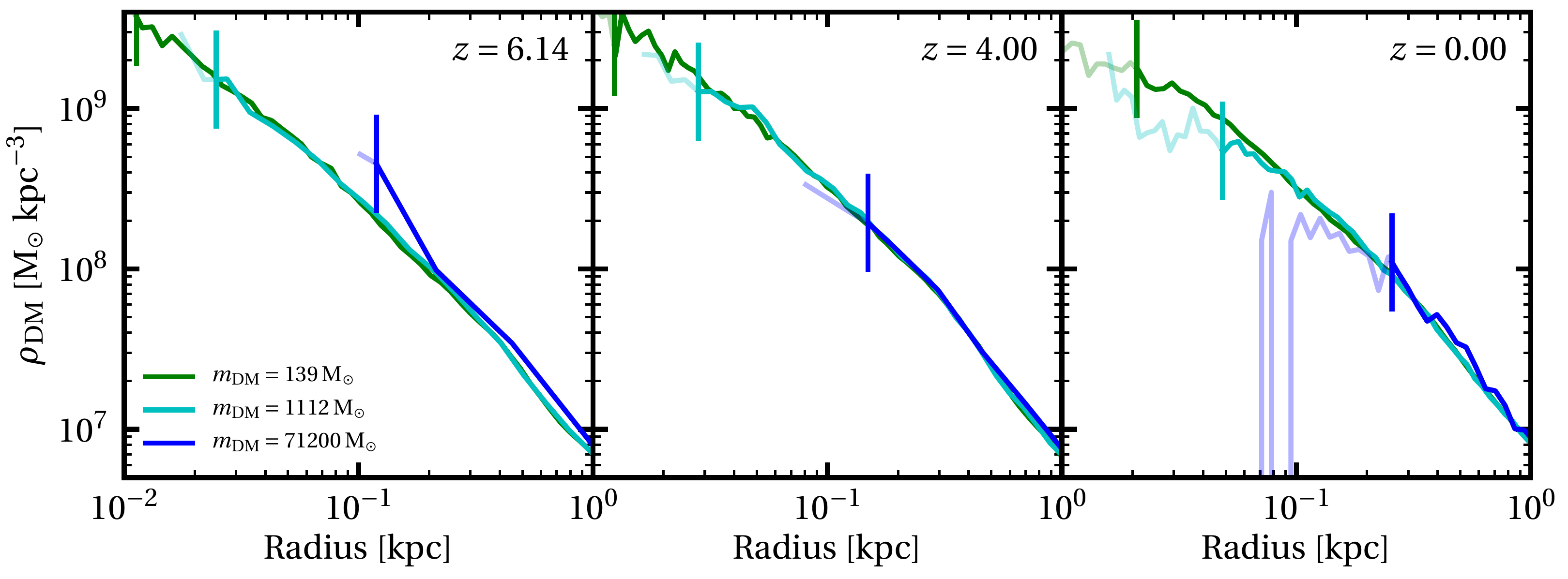}
\caption{The evolution of the 3D density profiles for Halo1459 DMO at three dark matter particle resolutions, as stated in the legend. The two lowest particle masses (green and cyan lines) correspond to the high resolution and fiducial resolutions, respectively (see Table \ref{tab:sims}). The fainter lines indicate the radii within which at least one relaxation time (estimated using equation \ref{relax.eq}) has passed, with the transition marked using an additional vertical line for clarity. In some cases, the density profiles are not plotted at smaller radii because there are insufficient particles.} \label{fig:resolution_study_eta}
\end{figure*}

The dark matter `particles' in collisionless cosmological simulations like {\sc ramses} are really `super-particles' that represent unresolved patches of the dark matter fluid. This approximation leads to overly large two-body relaxation that causes an artificial reduction in the inner dark matter density of a halo over time \citep[e.g.][]{2003MNRAS.338...14P,2004MNRAS.348..977D, 2004MNRAS.353..624D,2011EPJP..126...55D}.

The two-body relaxation timescale is given by \citet{1987gady.book.....B}:
\noindent\begin{equation}
t_{\rm relax} = \frac{N t_{\mathrm{orb}}}{16 \pi \log{\Lambda}},
\label{relax.eq}
\end{equation}
where $N$ is the number of particles within some radius $r$, $t_{\mathrm{orb}}$ is the orbital time of the system given by $t_{\mathrm{orb}} = 2 \pi \sqrt{R^3 / G M}$ and $\Lambda$ is the `Coulomb logarithm' given by $\Lambda = b_{\mathrm{max}} / b_{\mathrm{min}}$. The maximum impact parameter $b_{\mathrm{max}}$ and the minimum impact parameter $b_{\mathrm{min}}$ define the largest and smallest scales at which particles are expected to interact with each other. The relaxation time gives the time taken for the particle velocities to change by 90\degree, and can be considered as the time taken for a system to lose dynamical memory of its initial conditions. A large relaxation time is preferred because this implies two-body relaxation has a minimal influence on the particle kinematics. For standard $N$-body methods, the simplest way to increase the relaxation time is by increasing the number of dark matter particles, $N$, that sample the system \citep[e.g.][]{2011EPJP..126...55D}. \par

There is some debate in the literature over the best choices for the impact parameters $b_{\mathrm{max}}$ and $b_{\mathrm{min}}$. Here, we define $b_{\mathrm{max}}$ to be the total virial size of the system $r_{200}$, and $b_{\mathrm{min}}$ to be the side length of the highest resolution grid cell in {\sc ramses}. This is the approximate distance above which Newtonian gravity is recovered. Although multiple resolutions of grid cell are used throughout our simulated haloes, the central regions that are of interest here are predominantly at the highest resolution. \par

We now perform a brief resolution study in order to determine at what radius numerical relaxation effects become important. In Fig. \ref{fig:resolution_study_eta}, we plot dark matter density profiles from a DM-only simulation at three different resolutions. These three resolutions each exhibit a different amount of central density flattening due to numerical relaxation, with the flattening becoming stronger with increasing particle mass and with time. At $z=0$, all three resolutions are converged beyond $\approx 300\,$pc, but begin to deviate from the expected NFW form within some critical radius.

We define the `relaxation radius', $r_{\rm relax}$, to be the radius at which the relaxation time (equation \ref{relax.eq}) for the enclosed dark matter particles is equal to the simulation age for any particular simulation output: 
\noindent\begin{equation}
    \left ( \frac{r_{\rm relax}}{\text{kpc}} \right ) = \eta \left ( \frac{t_{\rm sim}}{\text{Gyr}} \right ) ^\alpha {\left ( \frac{M(r_{<\rm relax})}{\text{M}_{\odot}} \right )}^{-\frac{1}{3}} \left ( \frac{\left \langle m \right \rangle}{\text{M}_\odot} \right )^\frac{2}{3}
    \label{r_relax.eq}
\end{equation}
where we have substituted $t_{\rm relax}$ for the total run time of the simulation, $\left \langle m \right \rangle$ is the mean particle mass and $\eta$ and $\alpha$ are fitting parameters. From the data in Figure \ref{fig:resolution_study_eta}, we find $\alpha = 1/3$ and $\eta = 64 G \log{\Lambda} / 100^2$. The mass within the relaxation radius, $M(r_{\rm relax})$, is calculated directly from our simulation data and so equation \ref{r_relax.eq} can be solved numerically to find $r_{\rm relax}$. This is then used to predict the resolution limit for our EDGE simulations in this paper. The relaxation radii, calculated in this way, are marked on Figure \ref{fig:resolution_study_eta} by the vertical lines. Notice that the dark matter density profiles are shallower leftwards of $r_{\rm relax}$ in the lower resolution simulations as compared to the higher resolution simulations.

\section{Simulation convergence} \label{appendix:c}

\begin{figure}
\centering
\setlength\tabcolsep{2pt}%
\includegraphics[ trim={0cm 0cm 0cm 0cm}, clip, width=\columnwidth, keepaspectratio]{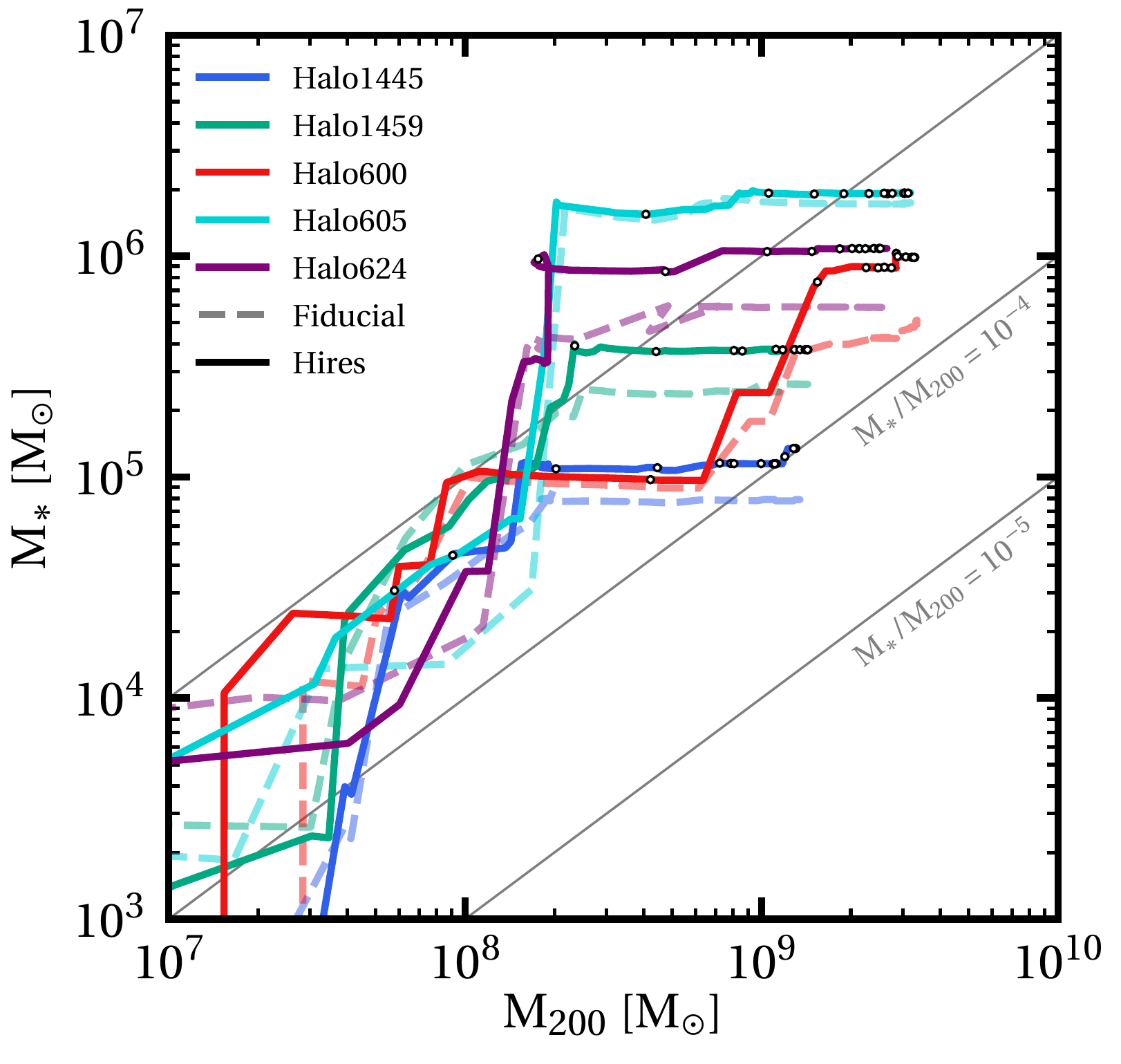}
\caption{The stellar mass-halo mass relation through time for our main simulation suite, where we compare our fiducial resolution simulations (faint dashed lines) to our high resolution simulations (solid lines). The black circles on the high resolution lines mark intervals of 1\,Gyr in time. The grey diagonal lines show constant ratios of $M{*}/M_{200 \rm c}$ in powers of ten, as marked}
\label{fig:SMHM_convergence}
\end{figure}

The dark matter particle mass resolution used in this paper is an order of magnitude smaller than used in the fiducial EDGE simulations \citep{2019ApJ...886L...3R, 2020MNRAS.491.1656A, 2020MNRAS.497.1508R, 2020MNRAS.tmp.3431P}. In this Appendix, we perform a convergence study to determine how our results are impacted by dark matter resolution. As previously in this paper, we distinguish between the lower and higher resolution simulations by appending the labels `fiducial' and `hires', respectively, to their name. \par

In Figure \ref{fig:SMHM_convergence}, we plot the total stellar mass within $r_{200 \rm c}$ versus $M_{200 \rm c}$ as a function of time for the fiducial (faint dashed) and hires (opaque solid) EDGE simulations. We find good convergence in shape of the general evolution of each simulated dwarf. The higher resolution simulations form systematically more stars (as first noted in \citealt{2020MNRAS.491.1656A}). However, the final stellar masses typically agree within $\sim 30$\% and at worst differ by a factor of $\sim$two (for Halo600). This is within the expected uncertainties due to modelling galaxy formation \citep{2020MNRAS.491.1656A}.

\begin{figure*}
\centering
\setlength\tabcolsep{2pt}%
\includegraphics[ trim={0cm 0cm 0cm 0cm}, clip, width=\linewidth, keepaspectratio]{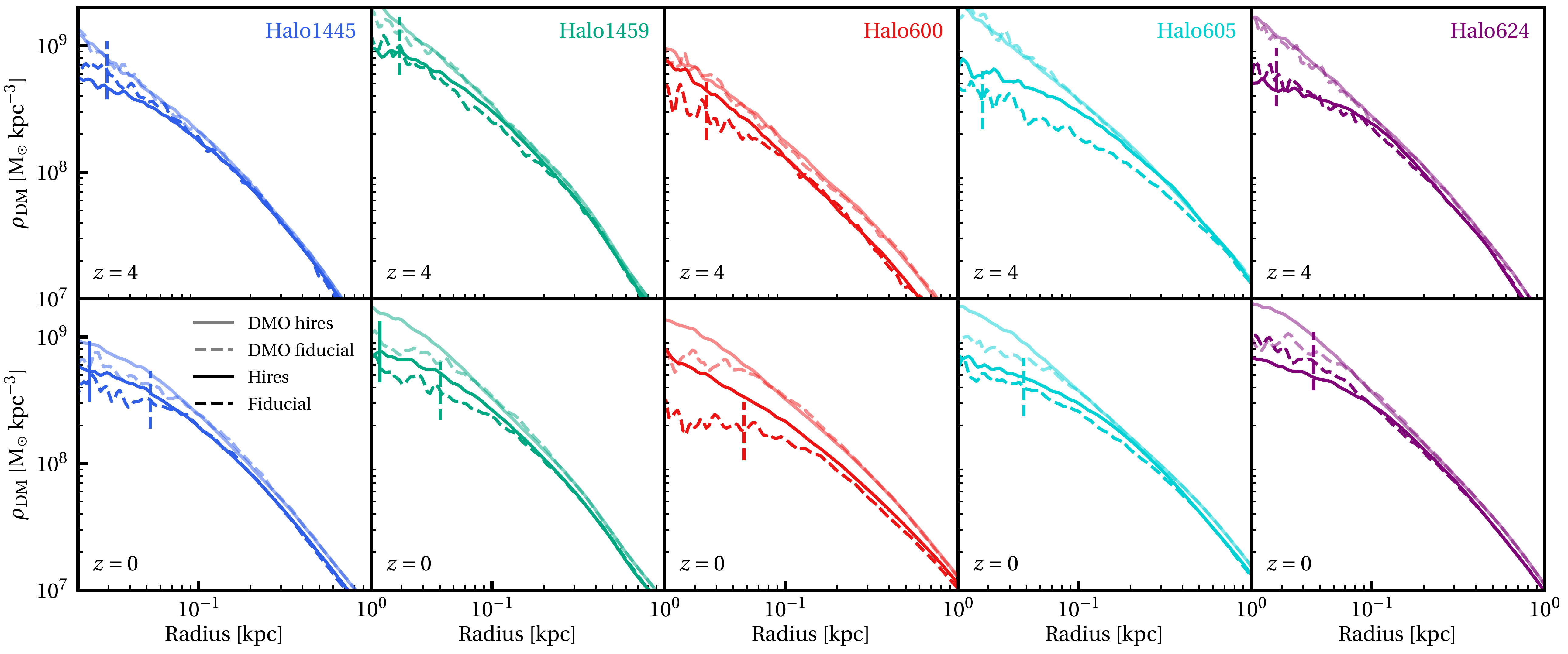}
\caption{The 3D dark matter density profiles of our main simulation suite, where we compare our DM-only simulations (faint lines) and baryonic simulations (opaque lines) at fiducial (dashed lines) and high (solid lines) resolution. The upper panels show this comparison at $z=4$, by which time all haloes are quenched due to reionisation (some will later reignite their star formation). The lower panels show this comparison at $z=0$. The numerical relaxation radius, as defined in Appendix \ref{appendix:a}, is indicated by a short vertical line for the baryonic simulations. This limit is similar for the corresponding DMO simulations.}
\label{fig:density_convergence}
\end{figure*}

In Figure \ref{fig:density_convergence}, we compare the 3D dark matter density profiles for our main simulation suite at two redshifts, $z=4$ (top panels) and $z=0$ (bottom panels). The DMO simulations are marked by the faint lines, the baryonic simulations by the opaque lines. The fiducial and hires simulations are marked by dashed and solid lines, respectively.


Overall, there is good convergence between the fiducial and hires simulations. The DMO simulations are all converged for $r > r_{\rm relax}$ (vertical lines), and there is good agreement also for the baryonic simulations Halo1445, Halo1459 and Halo624, at both $z=4$ and $z=0$. However, two of our EDGE simulations show poor convergence. Halo600 fiducial substantially rejuvenates its star formation after $z=1$, forming 17\% of its final stellar mass after this time. This extended period of star formation is replaced by a singular burst in Halo600 hires. As a result, Halo600 fiducial undergoes significantly more dark matter heating that causes its final inner density profile to be substantially lower than Halo600 hires. Similarly, Halo605 fiducial forms a larger dark matter core than Halo605 hires at early times. However, this disparity is largely resolved before $z=0$ due to a late cuspy merger in Halo605 fiducial (see Section~\ref{core destruction} for a discussion of this cusp reintroduction mechanism). 

The above results highlight an important point. It is often stated in the literature that whether a galaxy will be cusped or cored is determined by the stellar mass-to halo mass ratio, $M_*/M_{200 \rm c}$ \citep[e.g.][]{2012ApJ...759L..42P,2014MNRAS.441.2986D,2019MNRAS.484.1401R}. This is true to leading order. However, it also matter {\it how} those stars form, as illustrated by Halo600 fiducial versus Halo600 hires. The latter actually forms more stars, but because at late time they form in a single burst, this leads to less dark matter heating. And, at least at the very edge of galaxy formation, it also matters what the merger history is. Halo605 fiducial goes on to partially lose its dark matter core due to a late cuspy merger.

\par


\bsp	
\label{lastpage}
\end{document}